\pdfoutput=1
\documentclass[iop]{emulateapj}

\slugcomment{}

\shorttitle{Star formation activities toward {\it l} = 171$\degr$.7--174$\degr$.1}
\shortauthors{L.~K. Dewangan et al.}
\begin{document}

\title{Filamentary structures and star formation activities in the sites S234, V582, and IRAS 05231+3512}
\author{L.~K. Dewangan\altaffilmark{1}, T. Baug\altaffilmark{2}, D.~K. Ojha\altaffilmark{3}, I. Zinchenko\altaffilmark{4}, and A. Luna\altaffilmark{5}}
\email{Email: lokeshd@prl.res.in}
\altaffiltext{1}{Physical Research Laboratory, Navrangpura, Ahmedabad - 380 009, India.}
\altaffiltext{2}{Kavli Institute for Astronomy and Astrophysics, Peking 
University, 5 Yiheyuan Road, Haidian District, Beijing 100871, P. R. China.}
\altaffiltext{3}{Department of Astronomy and Astrophysics, Tata Institute of Fundamental Research, Homi Bhabha Road, Mumbai 400 005, India.}
\altaffiltext{4}{Institute of Applied Physics of the Russian Academy of Sciences, 46 Ulyanov st., Nizhny Novgorod 603950, Russia.}
\altaffiltext{5}{Instituto Nacional de Astrof\'{\i}sica, \'{O}ptica y Electr\'{o}nica, Luis Enrique Erro \# 1, Tonantzintla, Puebla, M\'{e}xico C.P. 72840.}
\begin{abstract}
To investigate the physical processes, we present observational results of the sites S234, V582, and IRAS 05231+3512 situated toward {\it l} = 171$\degr$.7--174$\degr$.1. 
Based on the CO line data, we find that these sites are not physically connected, and contain at 
least one filament (with length $>$ 7 pc). 
The observed line masses ($M_{\rm line,obs}$) of the filaments associated with V582 and 
IRAS 05231+3512 are $\sim$37 and $\sim$28 M$_{\odot}$ pc$^{-1}$, respectively. 
These filaments are characterized as thermally supercritical, and harbor several clumps. 
Groups of infrared-excess sources 
and massive B-type stars are observed toward the filament containing V582, while a very little star formation (SF) activity is found around IRAS 05231+3512. Our results favour radial collapse scenario in the filaments harboring V582 and IRAS 05231+3512. In the site S234, two filaments (i.e. ns1 ($M_{\rm line,obs}$ $\sim$130 M$_{\odot}$ pc$^{-1}$) and ns2 ($M_{\rm line,obs}$ $\sim$45 M$_{\odot}$ pc$^{-1}$)) are identified as thermally supercritical. 
An extended temperature structure at 27-30~K surrounds a relatively cold ($\sim$19 K) $\sim$8.9 pc long filament ns1. At least four condensations (M$_{clump}$ $\sim$70--300 M$_{\odot}$) are seen in ns1, and are devoid of the GMRT 610 MHz radio emission. 
The filament ns2 hosting clumps is devoid of ongoing SF, and could be at an early stage of fragmentation.
An intense SF activity, having the SF efficiency $\sim$3.3\% and SF rate $\sim$40--20 M$_{\odot}$ Myr$^{-1}$ (for t$_{sf}$ $\sim$1--2 Myr), is observed in ns1. The feedback of massive stars in S234 seems to explain the observed SF in the filament ns1.
\end{abstract}
\section{Introduction}
\label{sec:intro}
In recent years, the infrared and sub-millimeter (mm) data have revealed a wealth of bubbles and filamentary structures in star-forming regions \citep[e.g.][]{churchwell06,myers09,andre10}, 
which are often associated with star-forming clumps, clusters of young stellar objects (YSOs), and massive OB stars ($\geq$ 8 M$_{\odot}$). 
Understanding the role of filaments in the formation of dense star-forming clumps 
and the feedback of OB stars in their vicinity are still open research topics in 
the area of star formation \citep[e.g.][]{zinnecker07,myers09,andre10,deharveng10,schneider12,tan14,baug15,dewangan15,dewangan16,dewangan16b,dewangan17a,dewangan17b,dewangan17c,dewangan17d,dewangan17e}. 
Furthermore, the processes concerning the filament fragmentation are not well understood. 

In this paper, we have chosen promising star-forming sites located toward 
{\it l} = 171$\degr$.7 -- 174$\degr$.1; {\it b} = $-$0$\degr$.6 -- 0$\degr$.52 (see Figure~\ref{sg1}a), 
which include some of the major sites such as Sh 2-234 (hereafter S234), V582 Aur (hereafter V582), 
and Sh 2-237 (hereafter S237). The sub-mm map at 250 $\mu$m shows 
extended structures in these sites (see Figure~\ref{sg1}a). 
The sites S234, V582, and S237 are situated at distances of 2.8 kpc \citep{marco16,kun17}, 
1.3 kpc \citep{kun17,abraham18}, and 2.3 kpc \citep{pandey13,dewangan17a}, respectively. 
The CO radial velocities of the sites S234, V582, and S237 were reported 
to be $-$13.4 km s$^{-1}$ \citep[e.g.][]{blitz82}, $-$10.5 km s$^{-1}$ \citep[e.g.][]{abraham18}, 
and $-$4.5 km s$^{-1}$ \citep[e.g.][]{dewangan17a}. 

V582 is characterized as an FU Ori-type young eruptive 
star \citep[see][and references therein]{kun17,abraham18}, and 
the progenitor of V582 outburst is proposed to be 
a low-mass T Tauri star \citep{abraham18}. \citet{kun17} 
studied optical, near-infrared (NIR), and mid-infrared (MIR) data for 
a wide-field environment of V582, and suggested that V582 might be 
associated with a dark cloud LDN 1516. They further suggested that star 
formation in the vicinity of V582 could be triggered by the radiation field 
of a few hot members of Aur OB1 association. 

The site S237 has an almost sphere-like shell morphology, 
and is powered by a radio spectral type of B0.5V star \citep{dewangan17a}. 
In the site S237, \citet{dewangan17a} investigated a cluster of YSOs and a 
massive clump at the intersection of filamentary features, and explained the 
star formation through the collisions of the filamentary features \citep[see Figure 14 in][]{dewangan17a}.  

The site S234 (IC 417) harbors a nebulous stream (or filamentary structure) 
and a young cluster ``Stock 8" \citep{jose08,marco16}. 
Previous works also reported the presence of a dust shell lying between the nebulous 
stream and Stock 8 \citep[see Figures 1 and 9 in][]{marco16}, 
which was also found to be associated with the ionized emission \citep[e.g.][]{jose08}. 
\citet{marco16} pointed out that the nebulous stream is not directly linked with Stock 8. 
They identified about 15 early-type massive stars ($>$B2V) around Stock 8 based 
on the photometric and spectroscopic observations. 
These authors also suggested that the site S234 is powered by a star LS V +34$\degr$23 (HD 35633), 
with a spectral type of O8 II(f). \citet{jose08} and \citet{marco16} reported a stream 
of embedded sources and an obscured cluster (i.e. CC 14) toward the nebulous stream, and suggested 
the ongoing star formation activity in the site S234. 
It was pointed out that the YSOs associated with the nebulous stream \citep[timescale $\sim$1 Myr;][]{jose08} are younger than 
Stock 8 \citep[timescale $\sim$4--6 Myr;][]{marco16}. 
It has also been discussed that the star formation activities in the nebulous stream and Stock 8 may be independent \citep[e.g.][]{jose08}. 
Hence, these earlier studies together indicate that the site S234 contains several OB stars, 
nebulous stream/filamentary feature, dust shell, and clusters of YSOs. 
However, the molecular velocity structures in these different features 
have not been investigated yet.  

Taken together, the interesting extended features reported toward {\it l} = 171$\degr$.7 -- 174$\degr$.1; {\it b} = $-$0$\degr$.6 -- 0$\degr$.52 
are very suitable for investigating the role of filaments in star formation and the feedback of OB stars, which are yet to be carried out. 
To our knowledge, there is still lacking a detailed study of the molecular gas, embedded clumps, 
and ionized gas toward the sites in the direction of {\it l} = 171$\degr$.7 -- 174$\degr$.1. 
Therefore, new and unexplored sub-mm images, radio continuum maps, and molecular line data are analyzed in this paper to understand 
the ongoing physical processes in the star-forming sites.  

We give our observations and data analysis procedures in Section~\ref{sec:obser}. 
The results are presented in Section~\ref{sec:data}. 
Finally, the conclusions are drawn in Section~\ref{sec:conc}, which also includes a short summary of results.
\section{Data and analysis}
\label{sec:obser}
\subsection{New Observations}
\subsubsection{Radio Continuum Observations}
We obtained new radio continuum observations at 610 MHz of the site S234. 
The data were observed with the Giant Metre-wave Radio Telescope (GMRT) facility 
on 2016 December 08 (Proposal Code: 31\_025; PI: L.~K. Dewangan). 
The radio data reduction was carried out using the AIPS software, in a similar manner as 
given in \citet{mallick12,mallick13}. 
A synthesized beam size and an rms noise of the final 610 MHz continuum map 
are 5\farcs6 $\times$ 4\farcs6 and 120 $\mu$Jy/beam, respectively.

It is expected, specifically at low frequencies, that a sufficient amount of background 
emission will increase the antenna temperature, while observing the source. 
Hence, we corrected the final GMRT map at 610 MHz for system temperature \citep[see][]{omar02,mallick12,mallick13,baug15}. 
More details about the adopted procedure of system temperature corrections 
can be found in \citet{mallick12,mallick13}. 
\subsection{Archival data sets}
The multi-wavelength data sets are also retrieved from different surveys 
(e.g. the NRAO VLA Sky Survey \citep[NVSS; $\lambda$ = 21 cm; resolution = 45$\arcsec$;][]{condon98}, 
the Five College Radio Astronomy Observatory (FCRAO) $^{12}$CO (1-0) and $^{13}$CO (1-0) line data \citep[$\lambda$ = 2.6, 2.7 mm; resolution $\sim$45$\arcsec$;][]{heyer98,brunt04}, 
the {\it Herschel} Infrared Galactic Plane Survey \citep[Hi-GAL; $\lambda$ = 70, 160, 250, 350, 500 $\mu$m; resolutions = 5$''$.8, 12$''$, 18$''$, 25$''$, 37$''$;][]{molinari10}, the Wide Field Infrared Survey Explorer \citep[WISE; $\lambda$ = 3.4, 4.6, 12, 22 $\mu$m; resolutions = 6$''$.1, 6$''$.4, 6$''$.5, 12$''$;][]{wright10}, 
the Warm-{\it Spitzer} IRAC 3.6 and 4.5 $\mu$m photometric data \citep[GLIMPSE360; $\lambda$ = 3.6, 4.5 $\mu$m; resolution = 2$''$;][]{whitney11}, 
the UKIRT NIR Galactic Plane Survey \citep[GPS; $\lambda$ = 1.25, 1.65, 2.2 $\mu$m; resolution $\sim$0$''$.8;][]{lawrence07}, the Two Micron All Sky Survey \citep[2MASS; $\lambda$ = 1.25, 1.65, 2.2 $\mu$m; resolution = 2$''$.5;][]{skrutskie06}, and Isaac Newton Telescope (INT) Photometric H$\alpha$ Survey of the Northern Galactic Plane \citep[IPHAS; $\lambda$ = 0.6563 $\mu$m; resolution $\sim$1$''$;][]{drew05}). 
One can learn more details of these data sets and their analysis processes in \citet{dewangan17a}.
\section{Results}
\label{sec:data}
\subsection{Extended features in the direction of {\it l} = 171$\degr$.7 -- 174$\degr$.1}
\label{subsec:eff}
\begin{figure*}
\epsscale{1}
\plotone{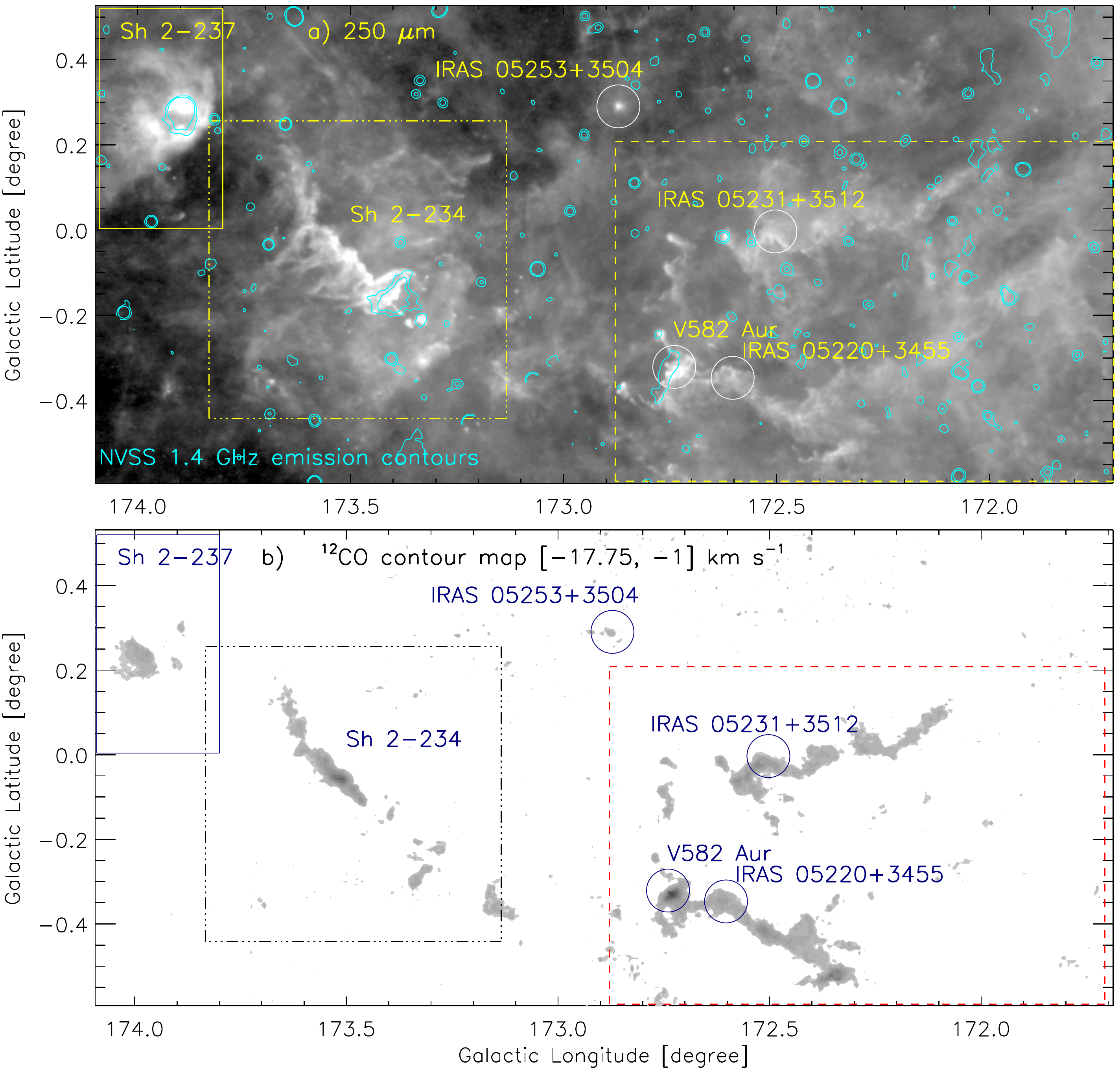}
\epsscale{1}
\plotone{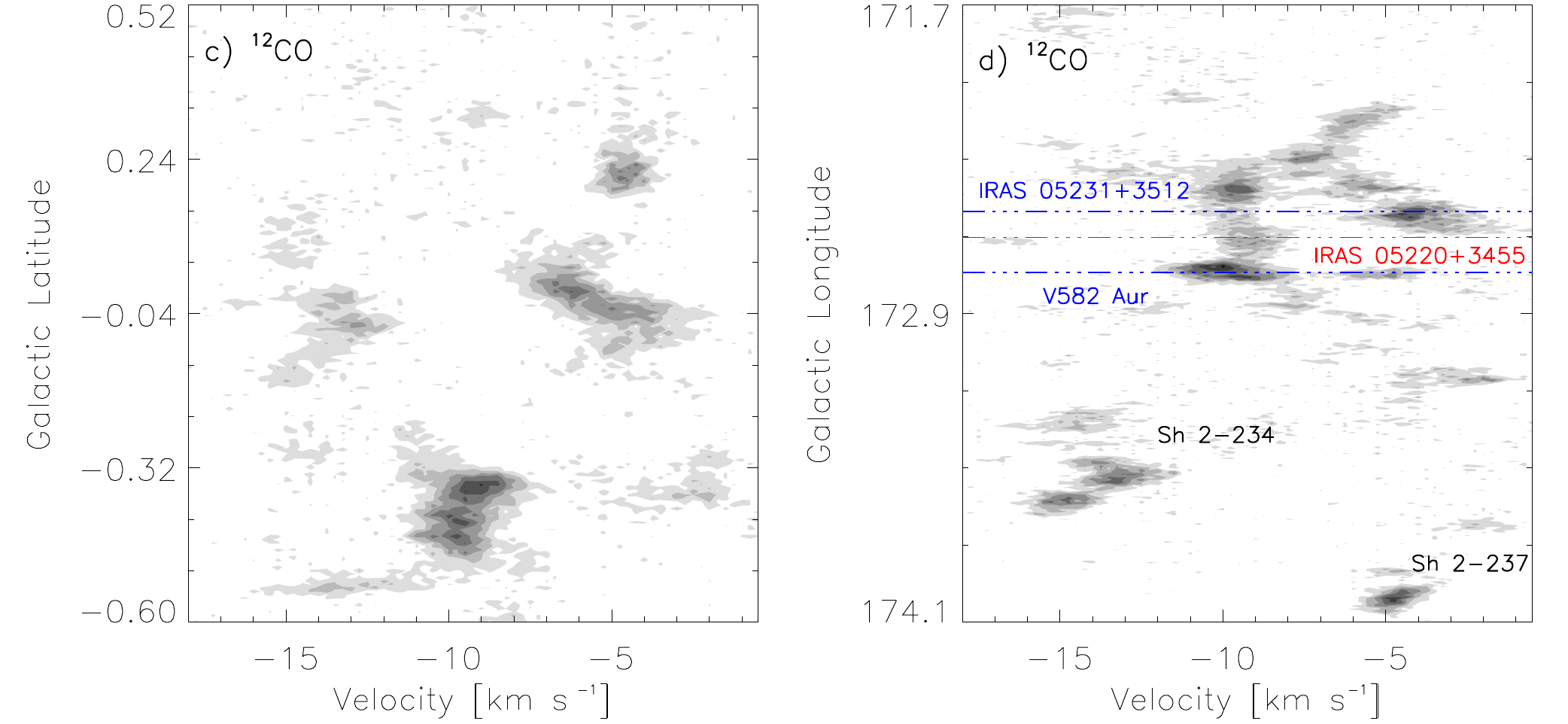}
\caption{a) {\it Herschel} 250 $\mu$m image of the $\sim$2$\degr$.4 $\times$ 1$\degr$.12 area toward {\it l} = 171$\degr$.7 -- 174$\degr$.1 and {\it b} = $-$0$\degr$.60 -- 0$\degr$.52, hosting the sites, Sh 2-237, Sh 2-234, IRAS 05253+3504, IRAS 05231+3512, IRAS 05220+3455, and V582 Aur. The image is also overlaid with the NVSS 1.4 GHz emission contours, and the contour levels are 0.45 mJy/beam $\times$ (3, 10). 
b) Integrated $^{12}$CO intensity map toward the field as shown in Figure~\ref{sg1}a.  
The $^{12}$CO emission is integrated over a velocity range of $-$17.75 to $-$1 km s$^{-1}$, and the contour levels are 81.2 K km s$^{-1}$ $\times$ (0.15, 0.2, 0.3, 0.4, 0.5, 0.6, 0.7, 0.8, 0.9, 0.98). 
c) Latitude-velocity plot of $^{12}$CO. The $^{12}$CO emission is integrated over 
the longitude from 171$\degr$.7 to 174$\degr$.1. 
d) Longitude-velocity plot of $^{12}$CO. The $^{12}$CO emission is integrated over the latitude from $-$0$\degr$.60 to 0$\degr$.52.
In the panels ``a" and ``b", the boxes highlight the vicinity of different star-forming sites.}
\label{sg1}
\end{figure*}

To explore the extended and embedded features toward {\it l} = 171$\degr$.7 -- 174$\degr$.1, 
we have employed the sub-mm image at 250 $\mu$m, radio continuum map at 1.4 GHz 
\citep[having a rms sensitivity  of $\sim$0.45 mJy/beam;][]{condon98}, and $^{12}$CO (J=1--0) line data 
\citep[having a velocity resolution of 0.25 km s$^{-1}$ and 
a rms sensitivity of 0.25 K;][]{heyer96,heyer98}. These data enable us to 
depict cold dust, ionized gas, and molecular gas in the sites.  
Figure~\ref{sg1}a displays the sub-mm image at 250 $\mu$m overlaid with the 1.4 GHz 
continuum emission contours. Figure~\ref{sg1}b shows the $^{12}$CO emission contour map at [$-$17.75, $-$1.0] km s$^{-1}$. 
We have labeled at least six objects (i.e. S237, S234, IRAS 05253+3504, IRAS 05231+3512, 
IRAS 05220+3455, and V582) in Figures~\ref{sg1}a and~\ref{sg1}b. 
The sub-mm image traces a sphere-like shell morphology of the site S237 and 
noticeable filamentary structures in the direction of S234, IRAS 05231+3512, 
IRAS 05220+3455, and V582. The extended ionized emission is observed toward the sites S237, S234, 
and V582 (see Figures~\ref{sg1}a). The filamentary morphologies of the molecular clouds 
associated with the sites S234, IRAS 05231+3512, IRAS 05220+3455, and V582 are revealed in Figure~\ref{sg1}b. 
To explore the molecular velocity structures, Figures~\ref{sg1}c and~\ref{sg1}d present the latitude-velocity 
and longitude-velocity plots of $^{12}$CO, respectively. 
Molecular gas in the sites S237, S234, IRAS 05231+3512, and V582 is traced in the velocity ranges of 
[$-$7, $-$2] km s$^{-1}$, [$-$17.75, $-$10] km s$^{-1}$, [$-$8.5, $-$1] km s$^{-1}$, 
and [$-$12.5, $-$7.5] km s$^{-1}$, respectively. 
These are different velocity ranges, indicating that these three sites 
do not belong to a single system, and are found at different distances (see Section~\ref{sec:intro}). 
The sites IRAS 05220+3455 and V582 are physically linked and embedded in the same filamentary feature, whereas the site IRAS 05231+3512 is embedded in another filamentary feature which has no physical association with the sites IRAS 05220+3455 and V582. 
Note that no observational study of the site IRAS 05231+3512 is available in the literature. 
The paper does not include the results of the site S237, which 
have been reported by \citet{dewangan17a}. 
A detailed study of the embedded filamentary features observed in the sites S234, IRAS 05231+3512, IRAS 05220+3455, 
and V582 is not available in the literature, and is presented in Sections~\ref{subsec:v582} and~\ref{subsec:s234}. 
\subsection{Physical environment and star formation in the sites V582 and IRAS 05231+3512}
\label{subsec:v582}
\begin{figure*}
\epsscale{0.67}
\plotone{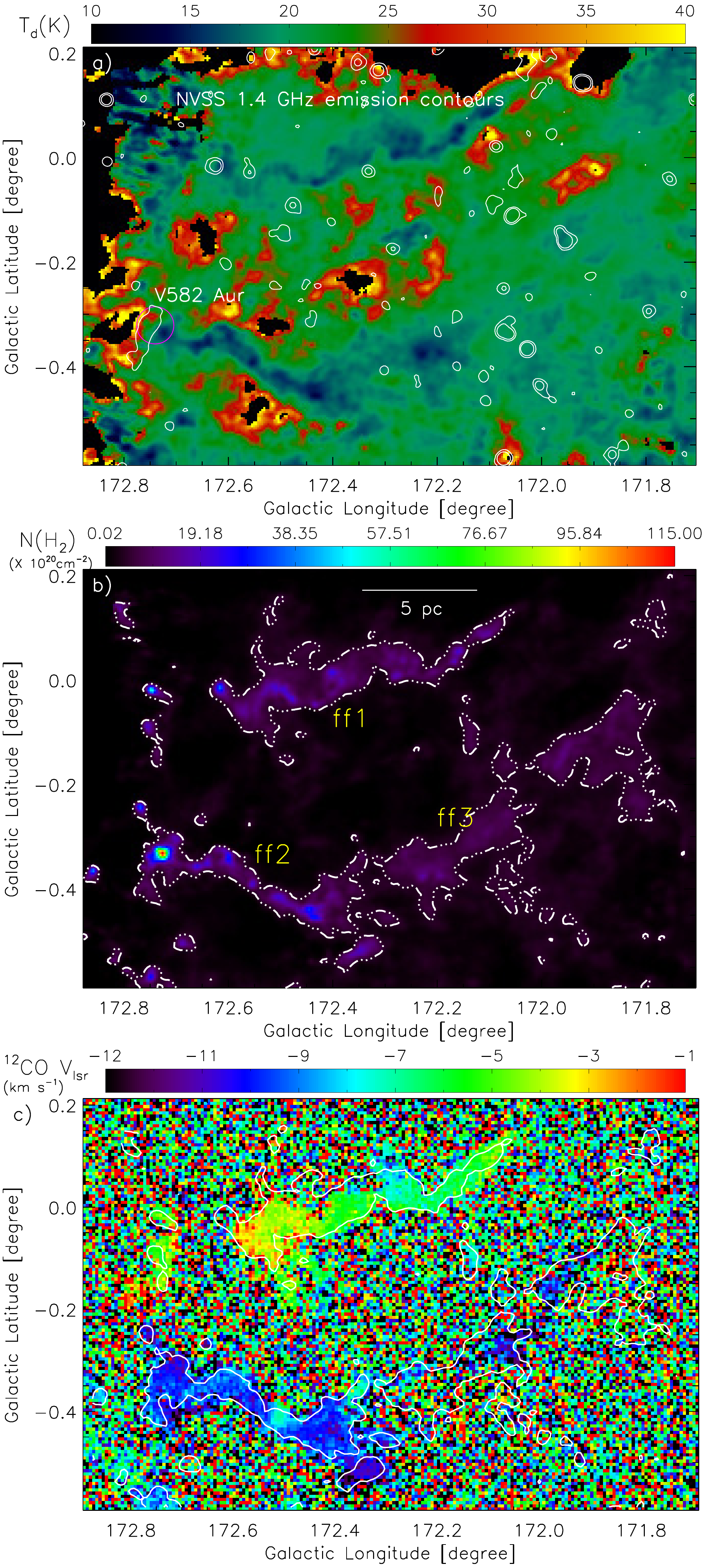}
\caption{a) {\it Herschel} temperature map of the $\sim$1$\degr$.17 $\times$ 0$\degr$.81 area of V582 Aur (see a dashed box in Figures~\ref{sg1}a and~\ref{sg1}b).
The NVSS 1.4 GHz emission contours with the levels of 0.45 mJy/beam $\times$ (3, 10) are also overlaid on the map. 
b) {\it Herschel} column density ($N(\mathrm H_2)$) map of the area as shown in Figure~\ref{sg2}a. 
The column density contour with a level of 4.5 $\times$ 10$^{20}$ cm$^{-2}$ is also overlaid on the map to trace different filamentary features. 
The scale bar corresponding to 5 pc (at a distance of 1.3 kpc) is shown in the panel. c) Velocity-field (moment 1) map of $^{12}$CO overlaid 
with the column density contour (in white). The contour is the same as in Figure~\ref{sg2}b.}
\label{sg2}
\end{figure*}

The temperature and column density maps (resolutions $\sim$37$''$) of the sites V582 and IRAS 05231+3512 
are presented 
in Figures~\ref{sg2}a and~\ref{sg2}b, respectively (see a dashed box in Figures~\ref{sg1}a and~\ref{sg1}b). 
Following the same analysis described in \citet{mallick15}, these maps are produced using the {\it Herschel} 160--500 $\mu$m images. The NVSS 1.4 GHz emission contours are also overlaid on the {\it Herschel} temperature map, tracing \
the ionized emission toward V582. 
A column density ($N(\mathrm H_2)$) contour with a level of 4.5 $\times$ 10$^{20}$ cm$^{-2}$ is 
also drawn in the {\it Herschel} column density map, allowing to infer filamentary features. 
At least three filamentary features are identified in the column density map, 
and are designated as ff1 (length $\sim$9.5 pc at a distance of 1.3 kpc), ff2 (length $\sim$12.7 pc at a distance of 1.3 kpc), 
and ff3 (length $\sim$6.9 pc at a distance of 1.3 kpc) (see Figure~\ref{sg2}b). 
The feature ff1 contains IRAS 05231+3512, while the sites IRAS 05220+3455 and V582 are 
associated with the feature ff2. The features ff1, ff2, and ff3 are well traced in temperatures of 14--18 K (see Figure~\ref{sg2}a).
Figure~\ref{sg2}c shows a first moment map of $^{12}$CO, revealing the distribution of mean velocities toward ff1, ff2, and ff3. 
The features ff2 and ff3 are traced in a velocity range of [$-$12.5, $-$7.5] km s$^{-1}$, while 
ff1 is depicted in velocities of [$-$8.5, $-$1] km s$^{-1}$.

\begin{figure*}
\epsscale{1}
\plotone{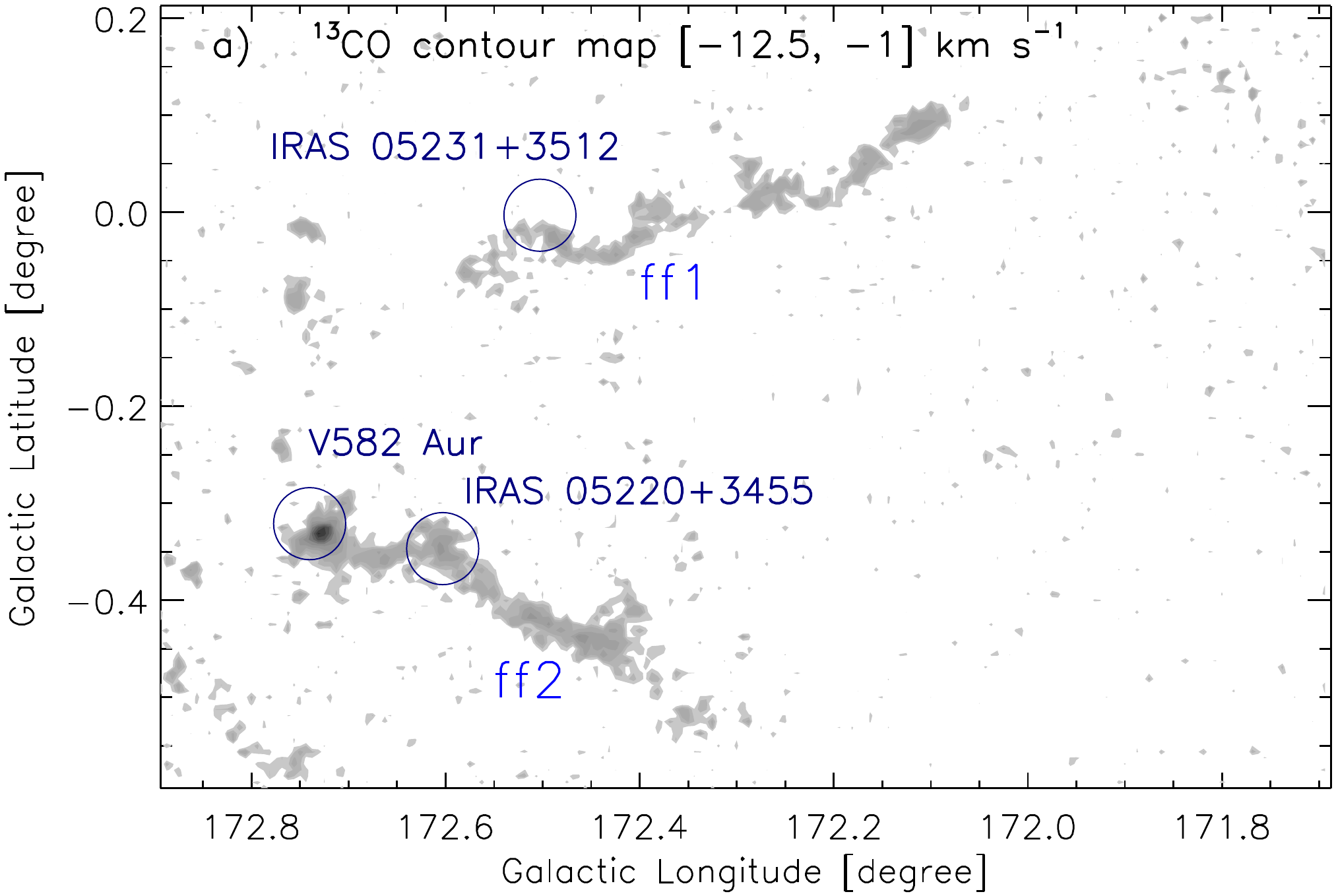}
\epsscale{1}
\plotone{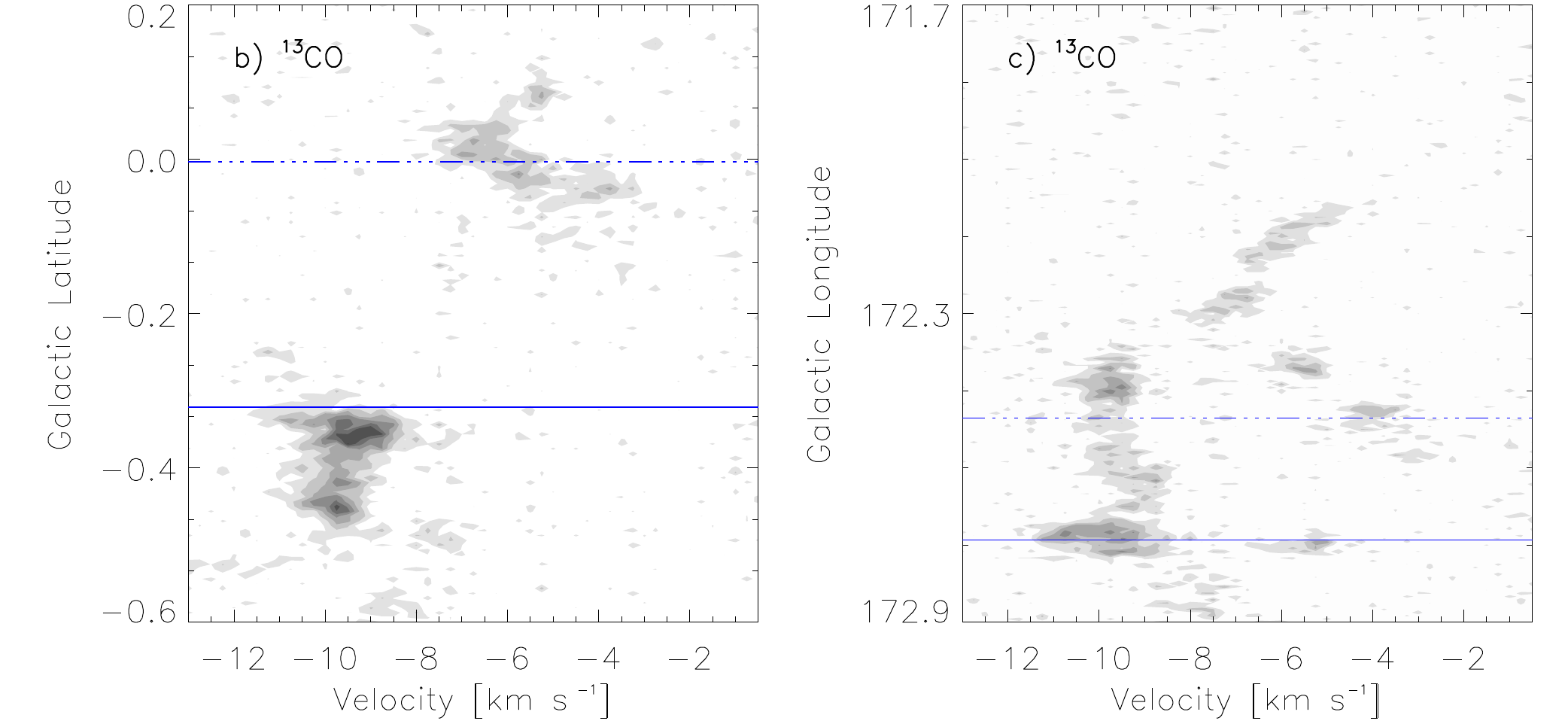}
\caption{Distribution of $^{13}$CO emission toward the area as shown in Figure~\ref{sg2}a.
a) Integrated $^{13}$CO intensity map in a velocity interval of [$-$12.5, $-$1] km s$^{-1}$. 
The contour levels are 22.4 K km s$^{-1}$ $\times$ (0.1, 0.15, 0.2, 0.3, 0.4, 0.5, 0.6, 0.7, 0.8, 0.9, 0.98). 
Big circles highlight the positions of IRAS 05231+3512, V582 Aur, and IRAS 05220+3455. 
b) Latitude-velocity plot of $^{13}$CO. 
The $^{13}$CO emission is integrated over the longitude from 171$\degr$.7 to 172$\degr$.9. 
c) Longitude-velocity plot of $^{13}$CO. 
The $^{13}$CO emission is integrated over the latitude from $-$0$\degr$.6 to 0$\degr$.2. 
In the panels ``b" and ``c", a dotted-dashed line (in blue) and a solid line (in blue) refer the positions of IRAS 05231+3512 and V582 Aur, respectively.}
\label{nng1}
\end{figure*}

We have computed a total mass of ff1, ff2, and ff3 to be 
$\sim$355, $\sim$362, and $\sim$182 M$_{\odot}$, respectively.
Here, the same procedures are adopted for computing the masses of {\it Herschel} features/clumps 
as given in \citet{mallick15}. The line masses or mass per unit lengths (i.e. $M_{\rm line,obs}$) of ff1, ff2, and ff3 are computed to be $\sim$37, $\sim$28, and $\sim$26 
M$_{\odot}$ pc$^{-1}$, respectively. Due to unknown inclination angles of the filamentary features, 
the observed line masses show upper limits. A critical line mass $M_{\rm line,crit}$ is 
calculated to be 16--24 M$_{\odot}$ pc$^{-1}$ at T = 10--15 K. Here, the conversion relation is equal to $\sim$16~M$_{\odot}$ pc$^{-1}$ $\times$ (T$_{gas}$/10 K) for 
a gas filament, assuming that the filament is an infinitely extended, self-gravitating, 
isothermal cylinder without magnetic support \citep[e.g.][]{ostriker64,inutsuka97,andre14}. If a condition $M_{\rm line,obs}$ $>$ $M_{\rm line,crit}$ is valid 
for a filament then it is a thermally supercritical filament, while a thermally subcritical filament 
is characterized with a condition $M_{\rm line,obs} < M_{\rm line,crit}$. Our analysis suggests that the 
features ff1, ff2, and ff3 are thermally supercritical filaments. 
The thermally supercritical filaments should be unstable to radial gravitational collapse and fragmentation \citep[e.g.][]{andre10}. 
In the direction of these filamentary features, several condensations, traced by column density peaks, are 
seen in the column density map (see Figure~\ref{sg2}b).
In Figure~\ref{nng1}a, we display the integrated $^{13}$CO \citep[having a velocity resolution of 0.25 km s$^{-1}$ and a rms sensitivity of 0.2 K;][]{heyer96,heyer98} intensity map, 
where the molecular emission is integrated over a velocity interval of [$-$12.5, $-$1] km s$^{-1}$. 
Molecular $^{13}$CO gas is observed only toward the features ff1 and ff2 in the molecular intensity map. 
One can keep in mind that the optical depth in $^{12}$CO is much higher than that in $^{13}$CO. 
Hence, the $^{13}$CO emission can trace relatively denser parts of a star-forming region compared to that of $^{12}$CO emission. 
Figures~\ref{nng1}b and~\ref{nng1}c show the latitude-velocity and longitude-velocity plots of $^{13}$CO, respectively.
These position-velocity maps also indicate the molecular gas in the features ff1 and ff2 at 
different velocity intervals as highlighted earlier. 

\begin{figure*}
\epsscale{1}
\plotone{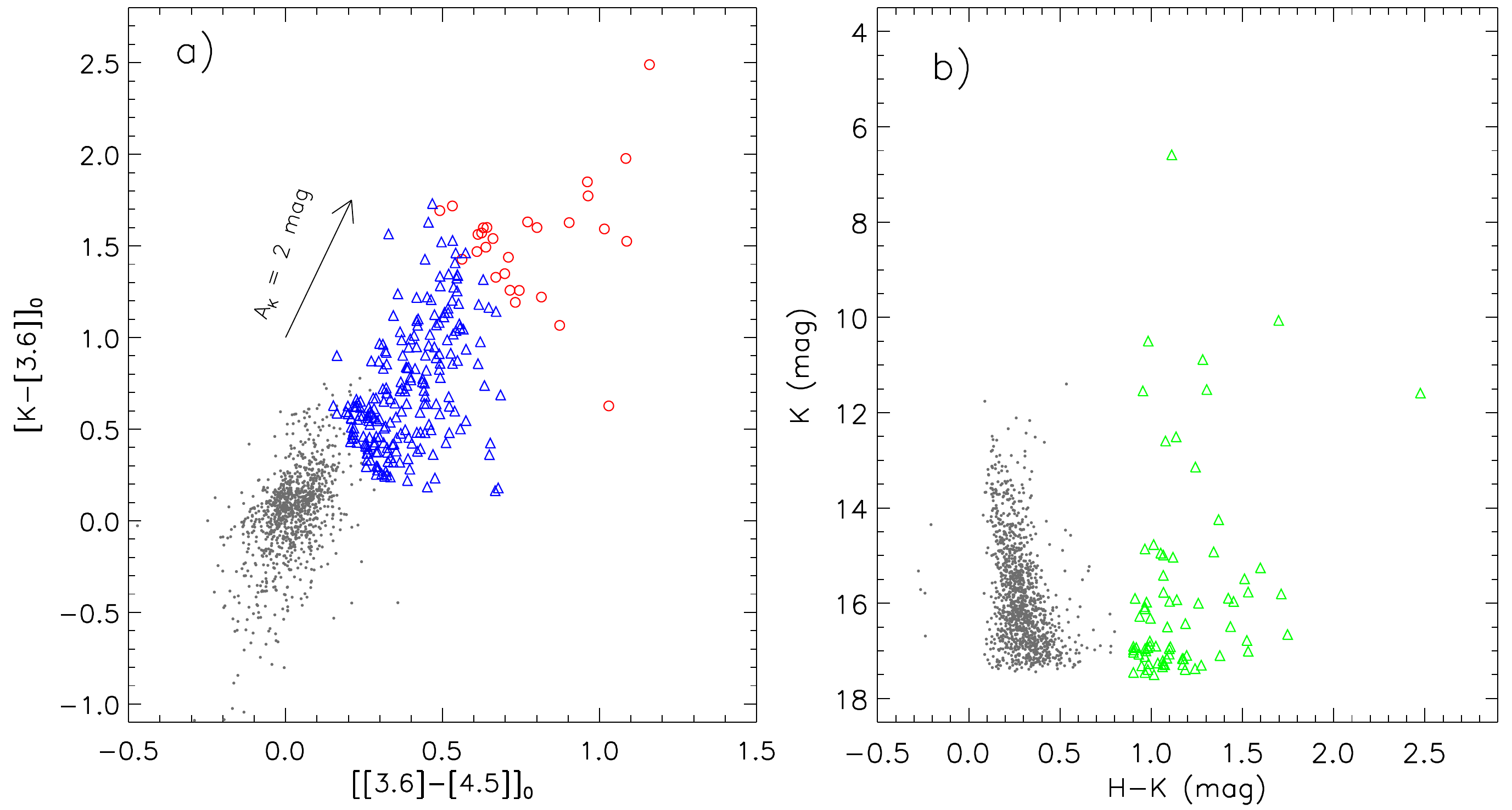}
\epsscale{1}
\plotone{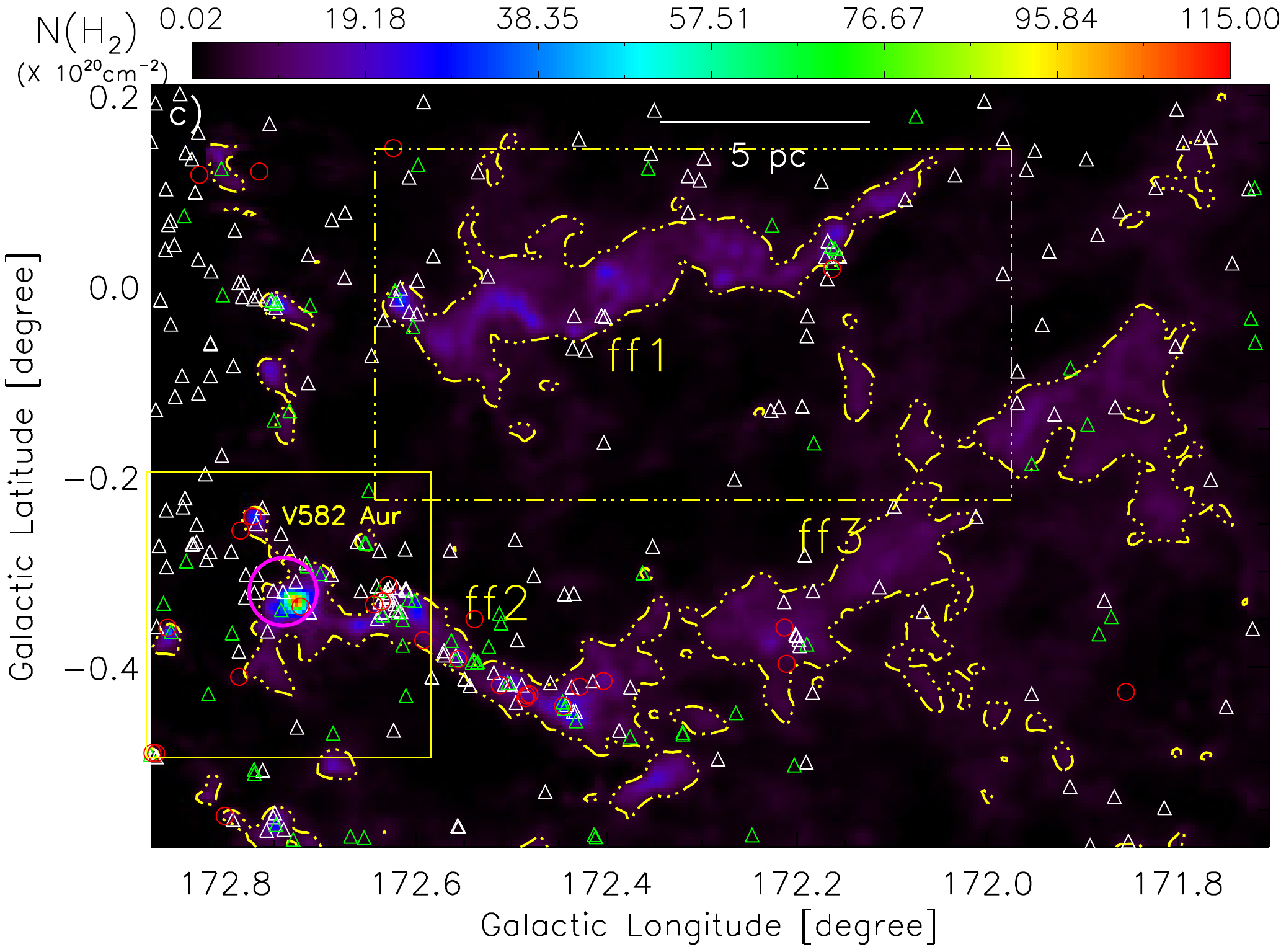}
\caption{a) Dereddened color-color ([K$-$[3.6]]$_{0}$ $vs$ [[3.6]$-$[4.5]]$_{0}$) plot 
of point-like objects toward the selected area around V582 Aur (see Figure~\ref{sg2}a). 
The extinction vector is also shown in the panel \citep[e.g.][]{flaherty07}. 
b) Color-magnitude (H$-$K/K) plot of point-like objects. 
c) Overlay of selected YSOs on the {\it Herschel} column density map.
In the panels ``a" and ``b", circles and triangles indicate Class~I and Class~II YSOs, respectively.
In the panels ``a" and ``b", the dots (in gray) present the stars with only photospheric emissions. 
We have randomly plotted only some of these stars in the top two panels. 
In the panel ``c", the YSOs highlighted with red circles and white triangles are selected based on the dereddened color-color scheme (see Figure~\ref{sg3}a), while green triangles refer the YSOs identified 
using the color-magnitude plot (see Figure~\ref{sg3}b).} 
\label{sg3}
\end{figure*}

Based on the {\it Spitzer}-GLIMPSE360, UKIDSS-GPS, and 2MASS photometric data, 
the dereddened color-color scheme (i.e. [K$-$[3.6]]$_{0}$ and [[3.6]$-$[4.5]]$_{0}$; see Figure~\ref{sg3}a) and 
the NIR color-magnitude scheme (i.e. H$-$K/K; see Figure~\ref{sg3}b) are utilized to identify the infrared-excess 
YSOs. More detailed descriptions of these schemes are given in \citet{dewangan17a} \citep[see also][]{gutermuth09}.
In the direction of our selected area around V582, color-color and color-magnitude schemes yield 243 (28 Class~I and 215 Class~II) YSOs, and 76 YSOs, respectively. 
Figure~\ref{sg3}c shows the spatial distribution of these YSOs overlaid on the column density map. 
The most of Class~I YSOs \citep[mean lifetime $\sim$0.44 Myr;][]{evans09} and Class~II YSOs \citep[mean lifetime $\sim$1--3 Myr;][]{evans09} are seen along the feature ff2 (hosting IRAS 05220+3455 and V582), revealing the noticeable star formation activity. 
However, there is a very little star formation activity found in the features ff1 and ff3 (see Figure~\ref{sg3}c).

\begin{figure*}
\epsscale{1}
\plotone{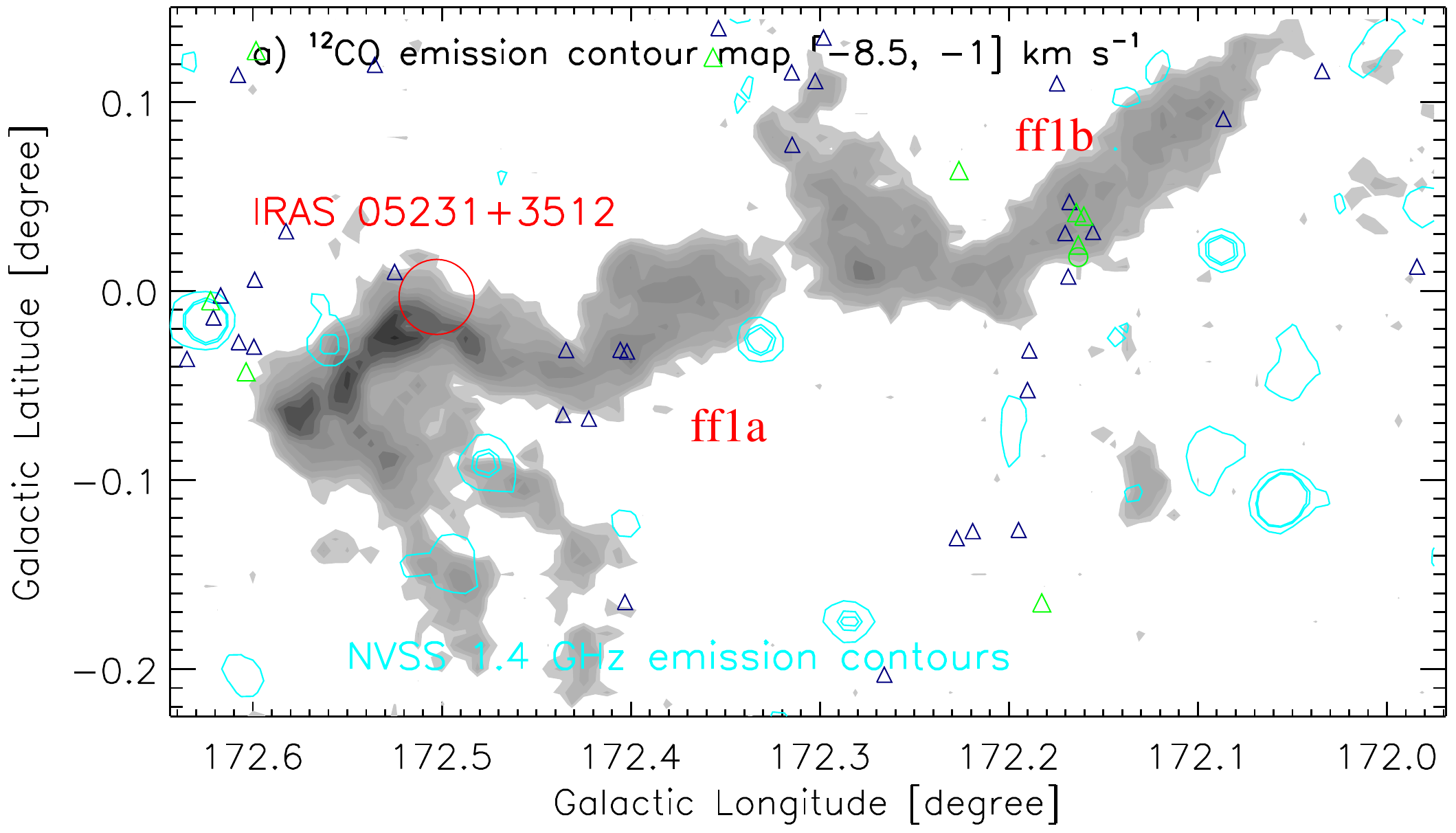}
\epsscale{0.7}
\plotone{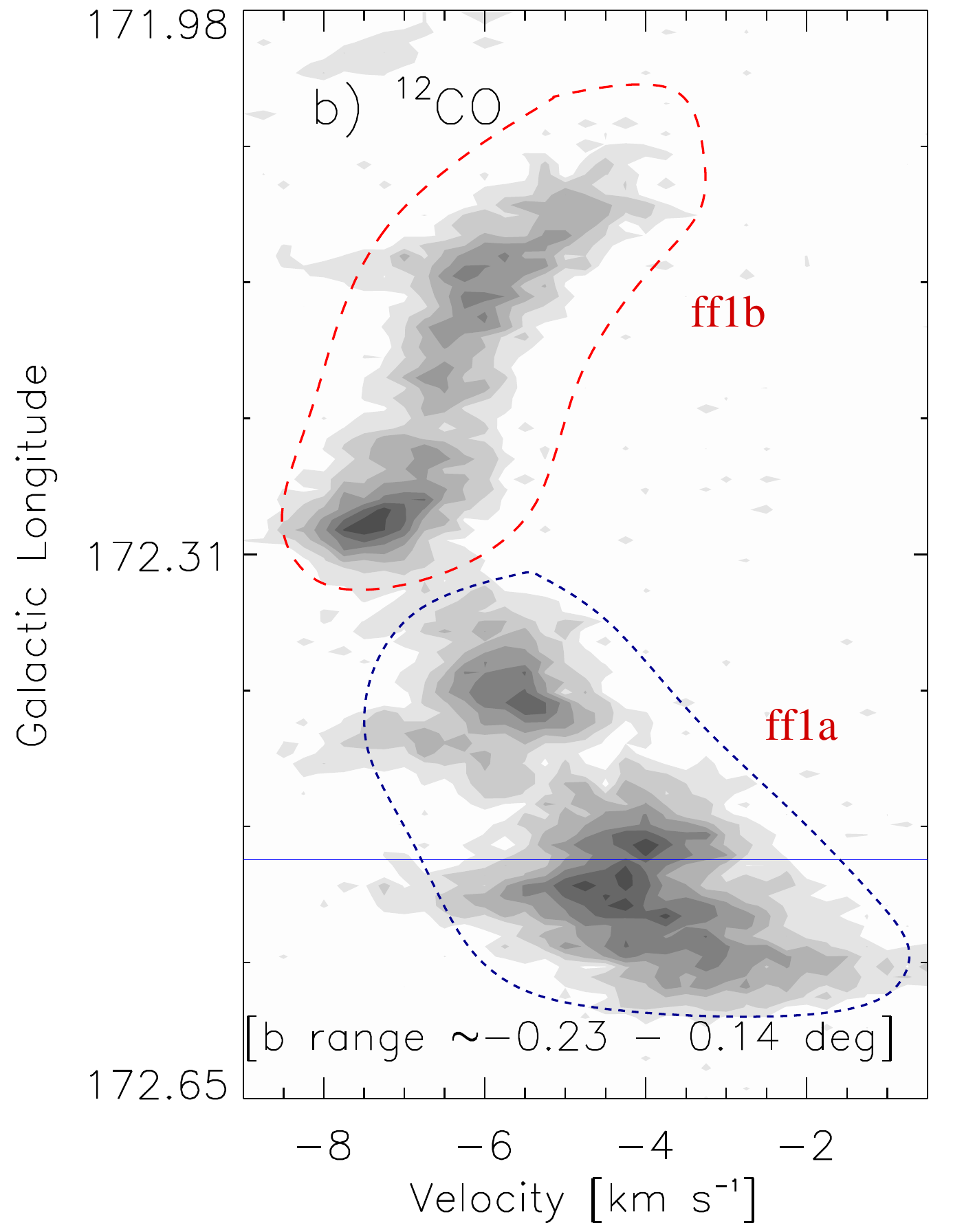}
\caption{A zoomed-in view of the feature ff1 (see a dotted-dashed box in Figure~\ref{sg3}c).
a) Integrated $^{12}$CO intensity map overlaid with the selected YSOs (see Figure~\ref{sg3}c)..  
The $^{12}$CO emission is integrated over a velocity range of $-$8.5 to $-$1 km s$^{-1}$, 
and the contour levels are 50.3 K km s$^{-1}$ $\times$ (0.1, 0.15, 0.2, 0.3, 0.4, 0.5, 0.6, 0.7, 0.8, 0.9, 0.98). 
A big circle highlights the position of IRAS 05231+3512. The NVSS contours (in cyan), with the levels of 0.45 mJy/beam $\times$ (3, 8, 10), are also overlaid on the molecular map. 
b) Longitude-velocity plot of $^{12}$CO. 
The $^{12}$CO emission is integrated over the latitude from $-$0$\degr$.23 to $-$0$\degr$.14. 
A solid line (in blue) indicates the position of IRAS 05231+3512. 
Two broken curves highlight two subcomponents (i.e. ff1a and ff1b) in the feature ff1.}
\label{vsg4}
\end{figure*}

In the direction of the feature ff1 containing IRAS 05231+3512, Figure~\ref{vsg4}a displays the integrated $^{12}$CO intensity map at [$-$8.5, $-$1] km s$^{-1}$ 
overlaid with the NVSS contours. There is no radio continuum emission observed toward IRAS 05231+3512. 
The selected YSOs are also overlaid on the molecular intensity map, where two distinct subcomponents (i.e. ff1a and ff1b) are highlighted.  
Figure~\ref{vsg4}b shows the longitude-velocity plot of $^{12}$CO toward the feature ff1, where the subcomponents ff1a and ff1b are indicated by broken curves. Molecular gas in the subcomponents is physically connected in the velocity space, confirming the feature ff1 as a single elongated structure. Furthermore, noticeable velocity gradient along each subcomponent is depicted 
in the position velocity plot (see Figure~\ref{vsg4}b). Our analysis suggests the existence of star-less clumps/condensations in the supercritical filament ff1.

\begin{figure*}
\epsscale{1}
\plotone{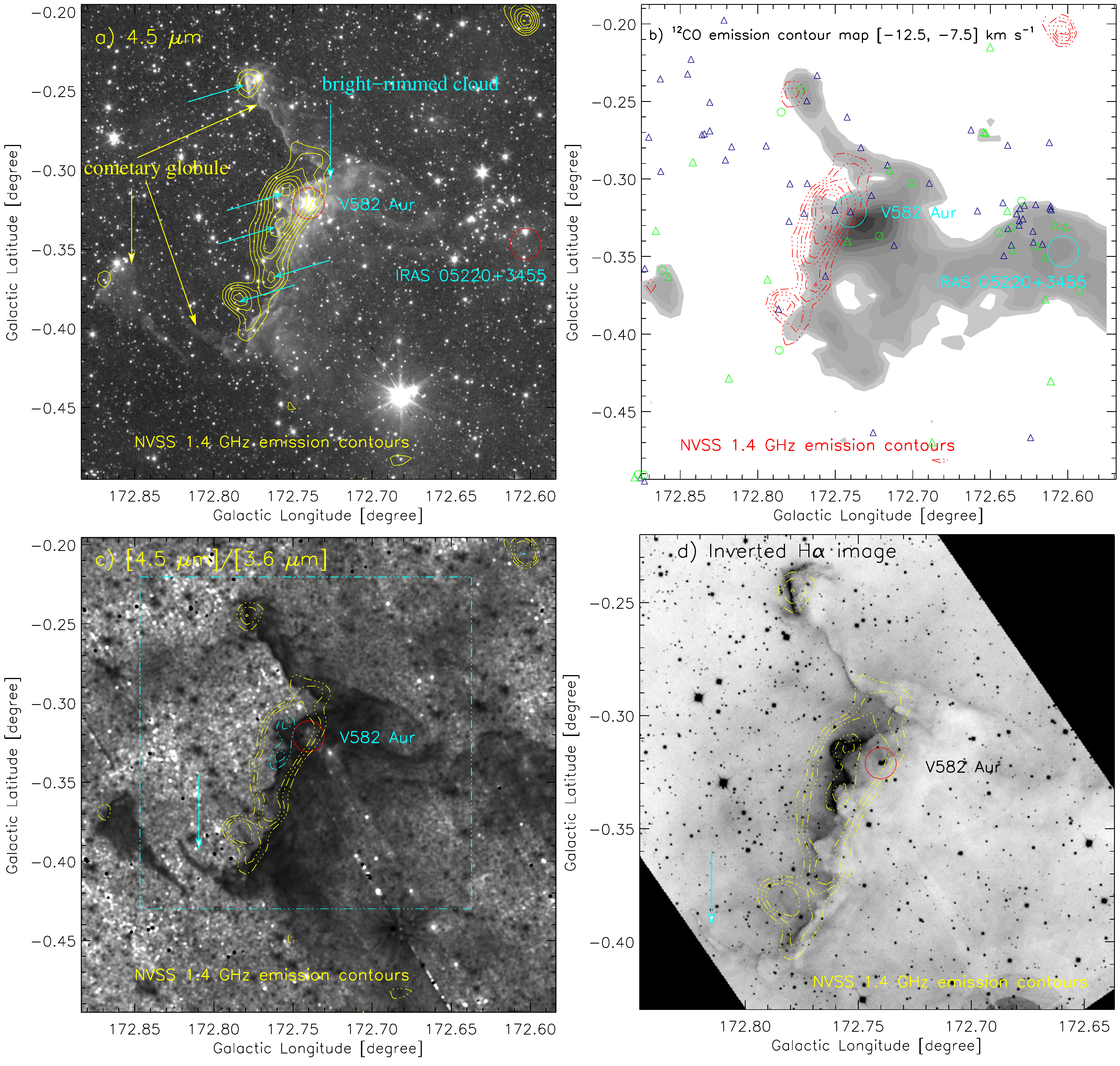}
\epsscale{1.04}
\plotone{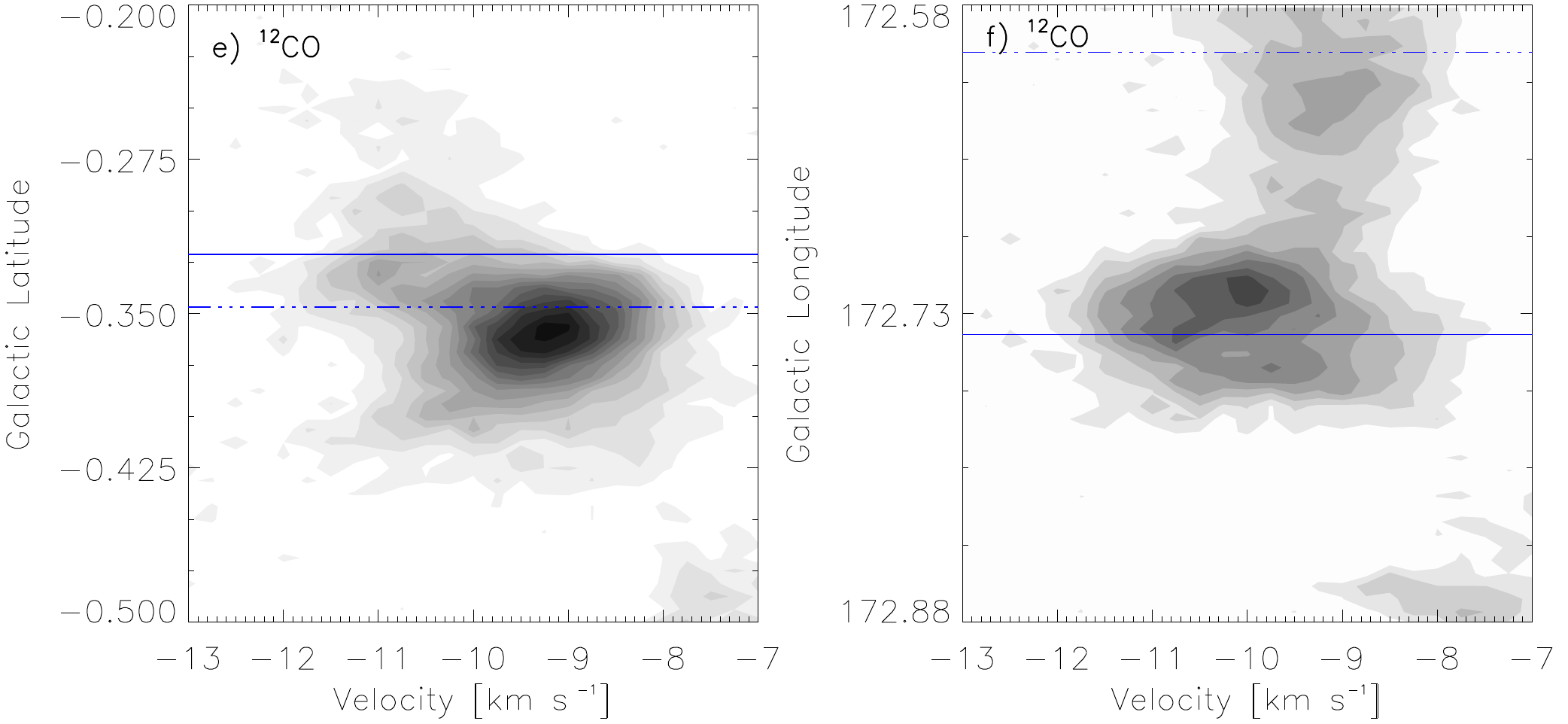}
\caption{A zoomed-in view of V582 Aur (see a solid box in Figure~\ref{sg3}c).
a) Overlay of the NVSS 1.4 GHz emission contours on the {\it Spitzer} 4.5 $\mu$m image. 
The contours are drawn with the levels of 0.45 mJy/beam $\times$ (3, 4, 5, 6, 7, 8, 9, 10).  
At least five radio peaks are indicated by arrows in the map. 
b) Integrated $^{12}$CO intensity map toward the field as shown in Figure~\ref{vsg3}a.  
The $^{12}$CO emission is integrated over a velocity range of $-$12.5 to $-$7.5 km s$^{-1}$, 
and the contour levels are 77.5 K km s$^{-1}$ $\times$ (0.1, 0.15, 0.2, 0.3, 0.4, 0.5, 0.6, 0.7, 0.8, 0.9, 0.98). 
The map is also overlaid with the NVSS contours (see Figure~\ref{vsg3}a) and the YSOs (see Figure~\ref{sg3}c). 
c) Overlay of the NVSS 1.4 GHz contours on the {\it Spitzer} ratio map of 4.5 $\mu$m/3.6 $\mu$m emission.
The ratio map is exposed to a Gaussian smoothing function with a width of 3 pixels. 
The dotted-dashed box (in cyan) encompasses the area shown in Figure~\ref{vsg3}d. 
d) Overlay of the NVSS 1.4 GHz contours on the IPHAS H$\alpha$ inverted gray-scale image. 
e) Latitude-velocity plot of $^{12}$CO. The $^{12}$CO emission is integrated over 
the longitude from 172$\degr$.58 to 172$\degr$.88. 
f) Longitude-velocity plot of $^{12}$CO. 
The $^{12}$CO emission is integrated over the latitude from $-$0$\degr$.50 to $-$0$\degr$.2.
In the panels ``a" and ``b", big circles highlight the positions of V582 Aur and IRAS 05220+3455. 
In the position-velocity plots, a solid and a broken lines (in blue) indicate the positions of V582 Aur and IRAS 05220+3455, respectively.}
\label{vsg3}
\end{figure*}

Using the {\it Spitzer} 4.5 $\mu$m image, Figure~\ref{vsg3}a shows a zoomed-in view of the feature ff2 (hosting IRAS 05220+3455 and V582), and reveals bright-rims and cometary 
globules in the local environment of V582. V582 is seen around one of the corners of the feature ff2. 
The image is also overlaid with the NVSS 1.4 GHz continuum emission contours. The ionized emission is mainly seen toward the bright-rims and the tips of the cometary globules. The NVSS 1.4 GHz map also shows at least five radio peaks (see arrows in Figure~\ref{vsg3}a). 
Based on the peak flux value of each radio peak, we find that each of the radio peaks is powered 
by a B3--B2 type star, indicating the presence of massive stars in the local vicinity of V582. However, the observed position of V582 does not coincide with any of the radio continuum peaks. 
For estimating the Lyman continuum photons and spectral type of each radio peak, we have utilized the same equation and procedures given in \citet{dewangan17a}. In the calculations, we considered a distance of 1.3 kpc and an electron temperature of 10$^{4}$ K. 
Here, we used the models of \citet{panagia73} to infer the spectral classes from the observed Lyman photons. 
Figure~\ref{vsg3}b displays the integrated $^{12}$CO intensity map overlaid with the NVSS contours. 
The $^{12}$CO emission is integrated over a velocity interval of [$-$12.5, $-$7.5] km s$^{-1}$. 
The selected YSOs are also marked in the molecular intensity map. 
In Figure~\ref{vsg3}b, star formation is evident toward the bright-rims and cometary globules. 
A group of YSOs is also found around IRAS 05220+3455, where the NVSS 1.4 GHz emission is absent. 
In Figures~\ref{vsg3}c and~\ref{vsg3}d, we display the overlay of the NVSS 1.4 GHz continuum emission on the {\it Spitzer}-IRAC ratio map of 4.5 $\mu$m/3.6 $\mu$m emission and the IPHAS H$\alpha$ inverted gray-scale image.
The details of the ratio map can be found in \citet{dewangan16}. 
Figure~\ref{vsg3}d displays the spatial association between the H$\alpha$ extended features and the radio continuum emission. The ratio map shows the extended black or dark gray regions due to the excess of 3.6 $\mu$m emission, Note that the {\it Spitzer} 3.6 $\mu$m band contains polycyclic aromatic 
hydrocarbon (PAH) emission at 3.3 $\mu$m. Hence, the ratio map traces photodissociation regions (or photon-dominated regions, or PDRs) around the bright-rims. 
It implies that the origin of the bright-rims and the cometary globules appears to be influenced 
by the feedback process (such as, radiation pressure, pressure-driven H\,{\sc ii} region, and stellar wind) of massive B type stars. 

Figures~\ref{vsg3}e and~\ref{vsg3}f show the latitude-velocity and longitude-velocity plots of $^{12}$CO, respectively. 
In the velocity space, a velocity spread is found in the direction of V582 (see Figure~\ref{vsg3}e), and 
the physical association of IRAS 05220+3455 and V582 is also evident (see Figure~\ref{vsg3}f). 
Together, the thermally supercritical filament ff2 is observed with the clusters of YSOs and massive stars.
\subsection{Physical environment and star formation in the site S234}
\label{subsec:s234}
\subsubsection{Multi-wavelength continuum images of S234}
\begin{figure*}
\epsscale{0.75}
\plotone{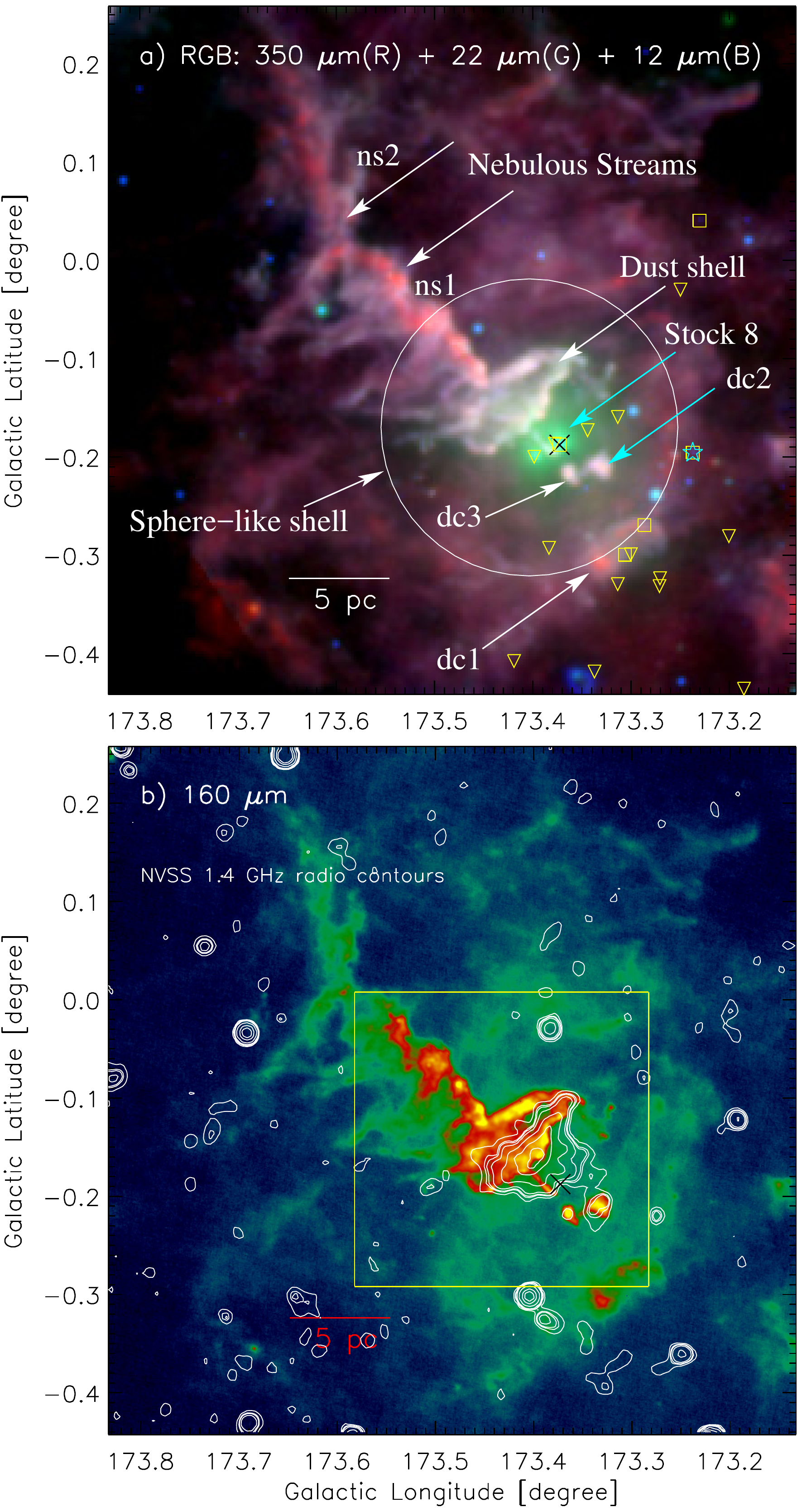}
\caption{Multi-wavelength images of the $\sim$0$\degr$.7 $\times$ 0$\degr$.7 area of the site S234 (see a dotted-dashed box in Figures~\ref{sg1}a and~\ref{sg1}b). 
a) A color-composite map of the site (red: {\it Herschel} 350 $\mu$m; green: {\it WISE} 22 $\mu$m; blue: {\it WISE} 12 $\mu$m images in log scale). 
A big circle (in white) shows a sphere-like shell structure. The yellow squares and upside down triangles (in yellow) show the positions of O and B-stars, respectively 
\citep[from][]{marco16}. The position of a star LS V +34$\degr$23, with a spectral type of O8 II(f) \citep{marco16}, 
is highlighted by a star. Arrows also indicate some observed features such as, ``Nebulous Streams (i.e. ns1 and ns2)", ``dust shell", ``dust condensations (i.e. dc1, dc2, and dc3)", and ``Stock 8". 
b) Overlay of the NVSS 1.4 GHz radio continuum emission (beam size $\sim$45$\arcsec$) on a false color {\it Herschel} 160 $\mu$m image.
The NVSS contour levels are 1, 1.7, 2, 2.8, 6, 13, and 20 mJy/beam. 
In each panel, the location of a star BD+34$\degr$1054 is marked by a big multiplication symbol. 
In both the panels, the scale bar corresponding to 5 pc (at a distance of 2.8 kpc) is 
shown in the bottom left corner.}
\label{sg4}
\end{figure*}

Figure~\ref{sg4}a shows a color-composite map (red: 350 $\mu$m; green: 22 $\mu$m, and blue: 12 $\mu$m) 
of the 0$\degr$.7 $\times$ 0$\degr$.7 area of the site (see a dotted-dashed box 
in Figures~\ref{sg1}a and~\ref{sg1}b). 
The positions of several OB stars \citep[from][]{marco16} are also marked in the composite map. 
At least two nebulous stream features, designated as ns1 and ns2, are prominently seen in the 350 $\mu$m image, 
while Stock 8 is more bright in the 12--22 $\mu$m images. 
The composite map also reveals a sphere-like shell 
structure (see a highlighted circle in Figure~\ref{sg4}a). 
The positions of a star LS V +34$\degr$23 \citep[spectral type: O8 II(f);][]{marco16} 
and a star BD+34$\degr$1054 \citep[LS V +34$\degr$29; spectral type: O9.7 IV;][]{marco16} are also highlighted in the composite map. 
The massive OB stars distributed around Stock 8 (see squares and upside down triangles 
in Figure~\ref{sg4}a) may explain the observed sphere-like shell structure through their feedback mechanism. A false color image at 160 $\mu$m of the site S234 is shown in Figure~\ref{sg4}b, and the map is also 
overlaid with the NVSS 1.4 GHz emission contours. 
A dust shell, where the ionized emission is observed, is remarkably seen in the MIR and 
sub-mm images (see Figures~\ref{sg4}a and~\ref{sg4}b). Additionally, at least three dust 
condensations, designated as dc1, dc2, and dc3, are also indicated in Figure~\ref{sg4}a. 
The massive OB stars are also seen toward the condensation dc1 (see squares and upside down triangles 
in Figure~\ref{sg4}a). 

\begin{figure*}
\epsscale{1.1}
\plotone{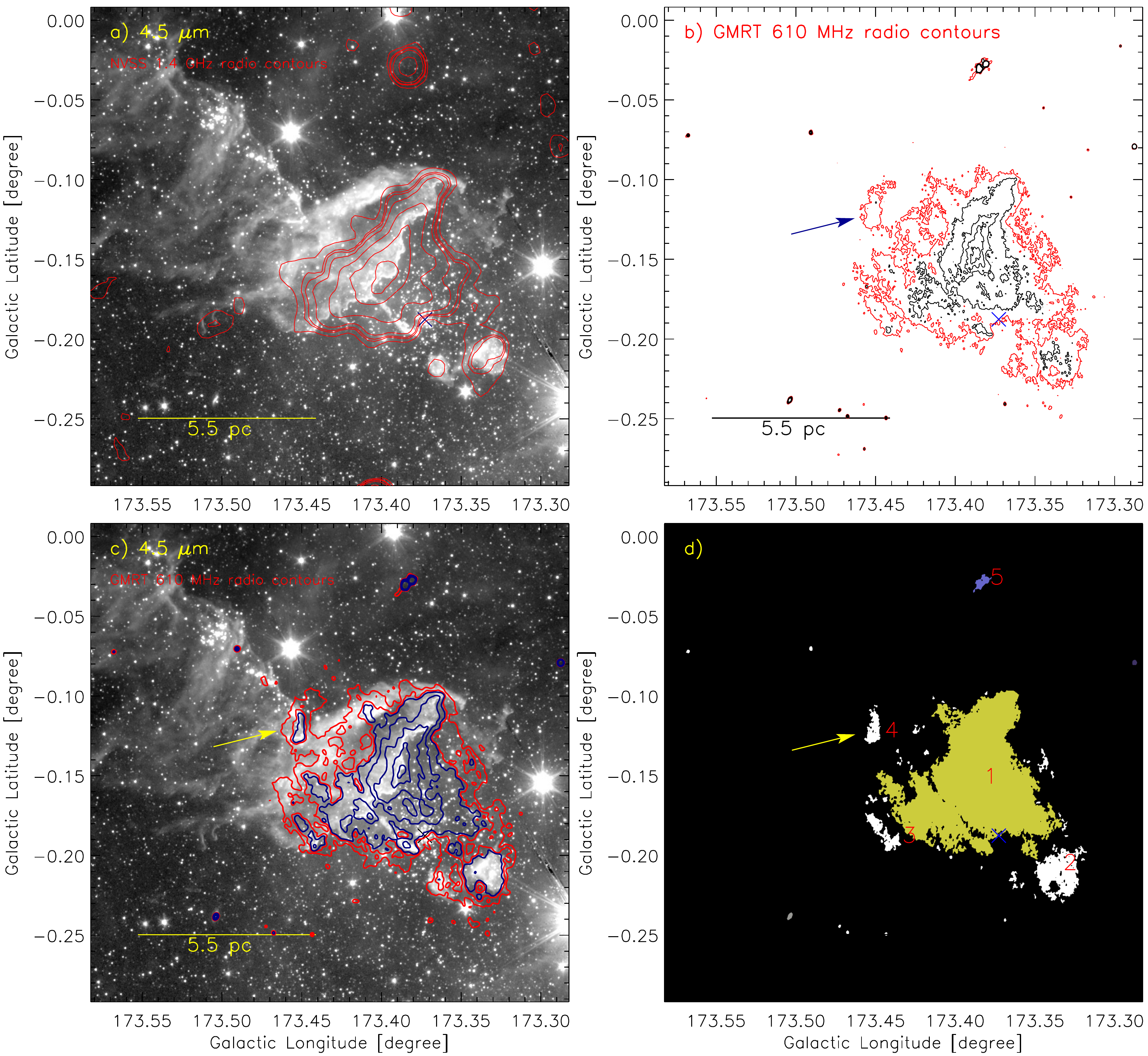}
\caption{Distribution of radio continuum emission toward the site S234 (see a solid box in Figure~\ref{sg4}b). 
a) Overlay of the NVSS 1.4 GHz emission contours (beam size $\sim$45$\arcsec$) on the {\it Spitzer} 4.5 $\mu$m image. 
The contour levels are the same as in Figure~\ref{sg4}b. 
b) GMRT radio continuum emission contours at 610 MHz (beam size $\sim$5$''$.6 $\times$ 4$''$.5; 1$\sigma$ $\sim$120 $\mu$Jy/beam). The contour (in red) is shown with a level of 3 $\times$ 120 $\mu$Jy/beam, while the other contours (in black) are drawn with the levels of 
120 $\mu$Jy/beam $\times$ (5, 8, 10, 12). c) Overlay of the GMRT 610 MHz emission contours on the {\it Spitzer} 4.5 $\mu$m image. 
The GMRT data are smoothened using a Gaussian function with radius of seven.
The contours (in red) are shown with the levels of 0.28 and 0.39 mJy/beam, while other 
contours (in navy) are drawn with the levels of 0.45, 0.65, 0.85, 1.2, and 1.5 mJy/beam. 
d) Clumpfind decomposition of the GMRT continuum emission. 
The extension of each identified ionized clump in the GMRT map is shown along with its corresponding ID and position (see also Table~\ref{tab1}).}
\label{sg5}
\end{figure*}

Using the {\it Spitzer} 4.5 $\mu$m image, a zoomed-in view of the dust shell is shown 
in Figure~\ref{sg5}a. The 4.5 $\mu$m image is also overlaid with the NVSS 1.4 GHz continuum emission. 
In Figure~\ref{sg5}b, we present new high-resolution GMRT radio continuum contours  
map at 610 MHz (resolution $\sim$5\farcs6 $\times$ 4\farcs6; rms $\sim$ 120 $\mu$Jy/beam). The GMRT radio contours are also overlaid on the {\it Spitzer} 4.5 $\mu$m image (see Figure~\ref{sg5}c), 
where the radio map is smoothened using a Gaussian function. 
Note that the GMRT map has better resolution and sensitivity compared to the NVSS data. 
Hence, the GMRT map provides more insight into the ionized features. 
The radio map at 610 MHz shows the distribution of the ionized gas toward the dust shell, 
Stock 8, and two dust condensations (i.e. dc2 and dc3). 
Furthermore, the ionized emission appears to be extended toward the nebulous 
stream ns1 in the GMRT map (see an arrow in Figures~\ref{sg5}b,~\ref{sg5}c, and~\ref{sg5}d), and is not detected in the NVSS map.
Table~\ref{tab1} lists physical parameters (i.e. position, radius, total flux, 
Lyman continuum photons, and spectral type) 
of five ionized clumps detected in the GMRT 610 continuum map. 
For estimating the Lyman continuum photons, we have used a distance of 2.8 kpc and an electron temperature of 10$^{4}$ K. 
The extension of each of the ionized clumps is also shown in Figure~\ref{sg5}d.  
All these ionized clumps are powered by massive B type stars (see Table~\ref{tab1}). 
To obtain the spectral class of each of the ionized clumps, the models of \citet{panagia73} are employed. 
\subsubsection{Molecular maps of S234}
\begin{figure*}
\epsscale{1.15}
\plotone{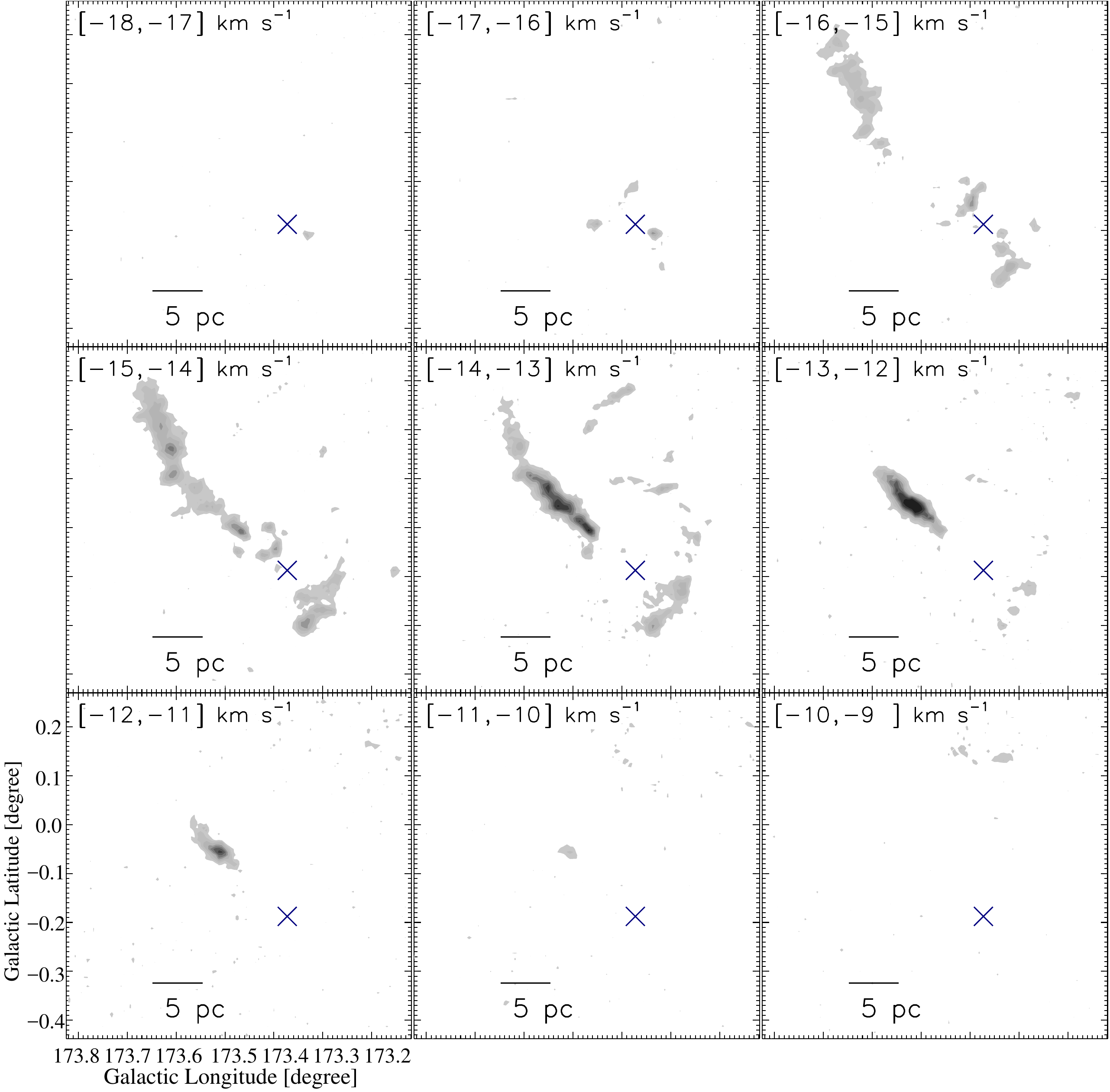}
\caption{Velocity channel contour maps of $^{12}$CO(J =1$-$0). 
The molecular emission is integrated over a velocity interval, which is marked in each panel (in km s$^{-1}$). 
The contour levels of $^{12}$CO emission are 2.5, 5, 8, 12, 15, 18, 21, and 25 K km s$^{-1}$. In each panel, the location of a star BD+34$\degr$1054 is marked by a big multiplication symbol.}
\label{sg6}
\end{figure*}

\begin{figure*}
\epsscale{0.58}
\plotone{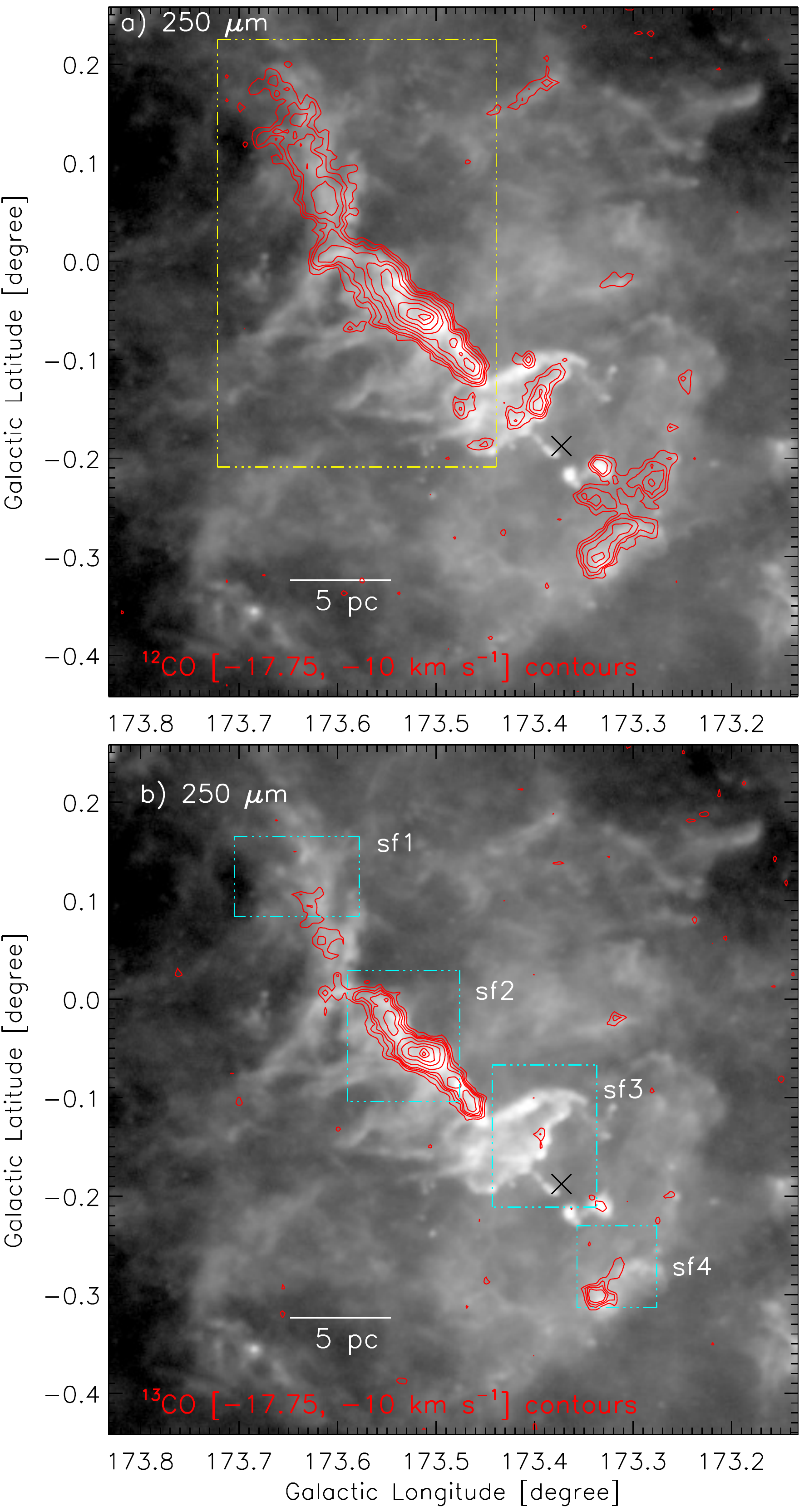}
\epsscale{0.56}
\plotone{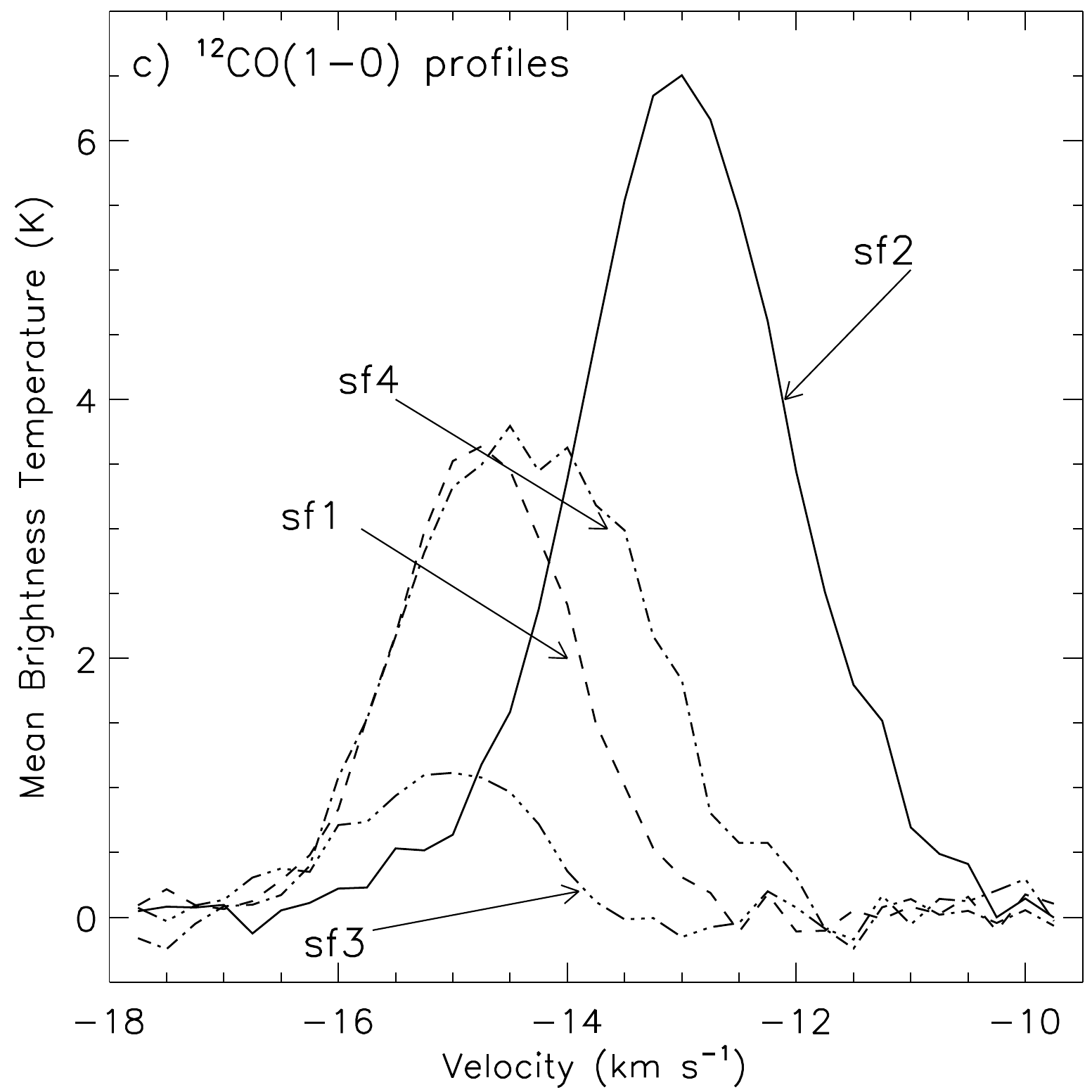}
\caption{Overlay of the molecular gas ($^{12}$CO (a) and $^{13}$CO (b)) on the {\it Herschel} image at 250 $\mu$m. The CO emission is integrated over a velocity range of $-$17.75 to $-$10 km s$^{-1}$. The contours of $^{12}$CO are shown with levels of 10, 15, 20, 30, 40, 55, 70, 85, and 95\% of the peak value (i.e. 70.8 K km s$^{-1}$), while the $^{13}$CO contours are 22 K km s$^{-1}$ $\times$ (0.13, 0.2, 0.25, 0.3, 0.4, 0.55, 0.7, 0.85, and 0.95). c) The average $^{12}$CO profiles in the direction of four small fields (i.e. sf1 to sf4; see corresponding boxes in 
Figure~\ref{sg7}b).}
\label{sg7}
\end{figure*}

Figure~\ref{sg6} displays the integrated FCRAO $^{12}$CO (J=1-0) 
velocity channel maps (at intervals of 1 km s$^{-1}$). Based on the gas distribution in different velocities, the maps confirm the existence of two 
nebulous streams in the site S234. 
Figures~\ref{sg7}a and~\ref{sg7}b show the integrated $^{12}$CO and $^{13}$CO emission contours overlaid on the 
{\it Herschel} 250 $\mu$m map, respectively. 
The CO emission is integrated over a velocity interval of [$-$17.75, $-$10] km s$^{-1}$. 
Molecular gas is not observed toward Stock 8. 
The $^{12}$CO emission is detected toward the nebulous streams, dust shell, and 
dust condensations, as highlighted in Figure~\ref{sg4}a, 
while the $^{13}$CO emission is prominently seen toward the features ns1 and dc1. 
It implies that the features ns1 and dc1 seem to be more denser regions in the site S234. 
Figure~\ref{sg7}c displays the average $^{12}$CO profiles toward four small fields (i.e. sf1, sf2, sf3, and sf4). 
The fields sf1, sf2, sf3, and sf4 are selected in the direction of the features ns2, ns1, dust shell, and dust condensation, respectively. 
In the profiles, the molecular velocity peaks toward the fields sf1, sf3, and sf4 are almost same (i.e. around $-$15 km s$^{-1}$), 
while the velocity peak in the direction of the field sf2 is seen around $-$13 km s$^{-1}$.

\begin{figure*}
\epsscale{1.1}
\plotone{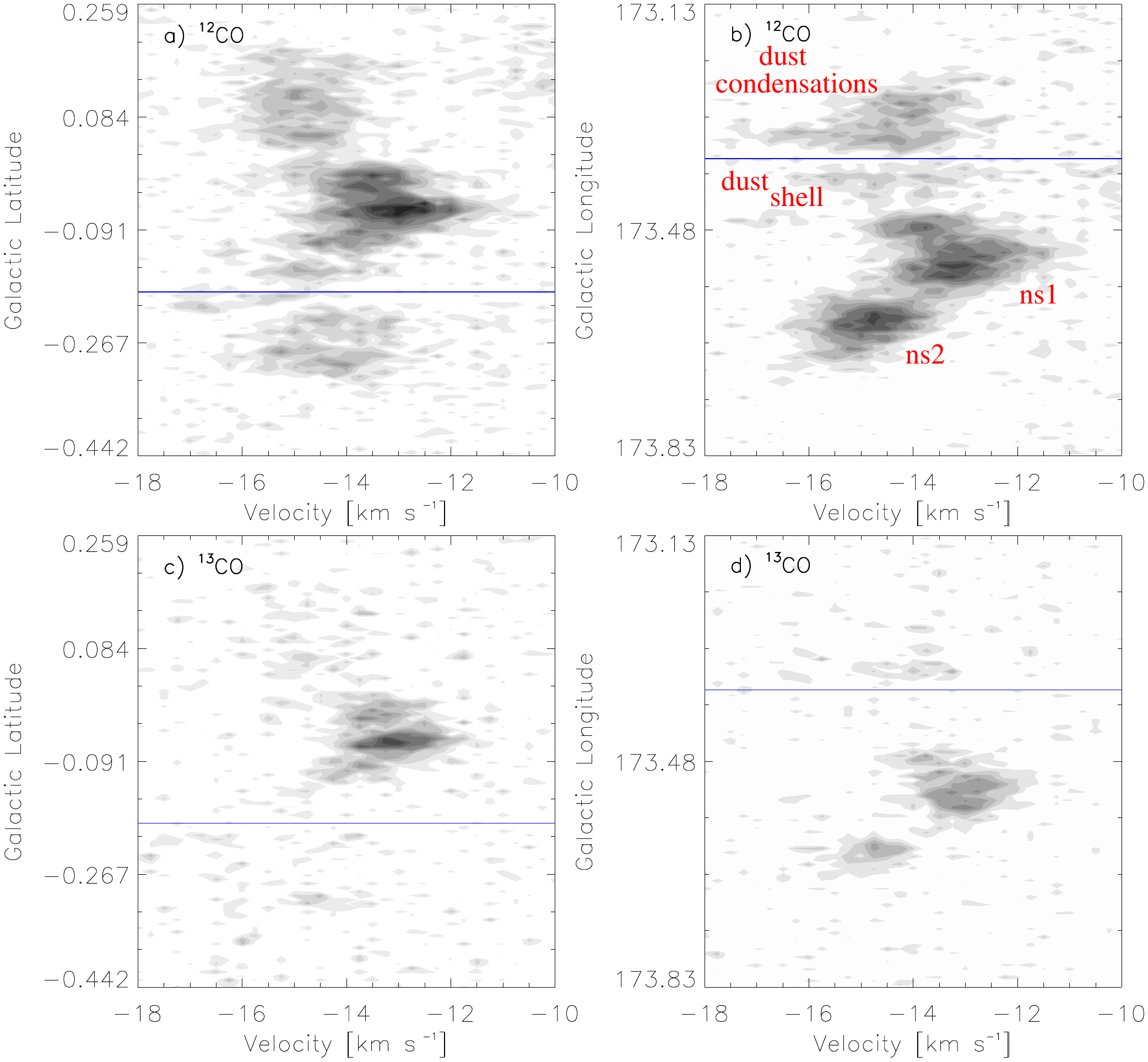}
\caption{Position-velocity plots toward the site S234. 
a) Latitude-velocity plot of $^{12}$CO.
b) Longitude-velocity plot of $^{12}$CO.
c) Latitude-velocity plot of $^{13}$CO.
d) Longitude-velocity plot of $^{13}$CO.
In both the left panels, the CO emission is integrated over 
the longitude from 173$\degr$.13 to 173$\degr$.83.
In both the right panels, the CO emission is integrated over 
the latitude from $-$0$\degr$.442 to 0$\degr$.259.
In each plot, a solid line (in blue) indicates the position of a star BD+34$\degr$1054 (see also Figure~\ref{sg7}a).}
\label{sg8}
\end{figure*}

\begin{figure*}
\epsscale{0.88}
\plotone{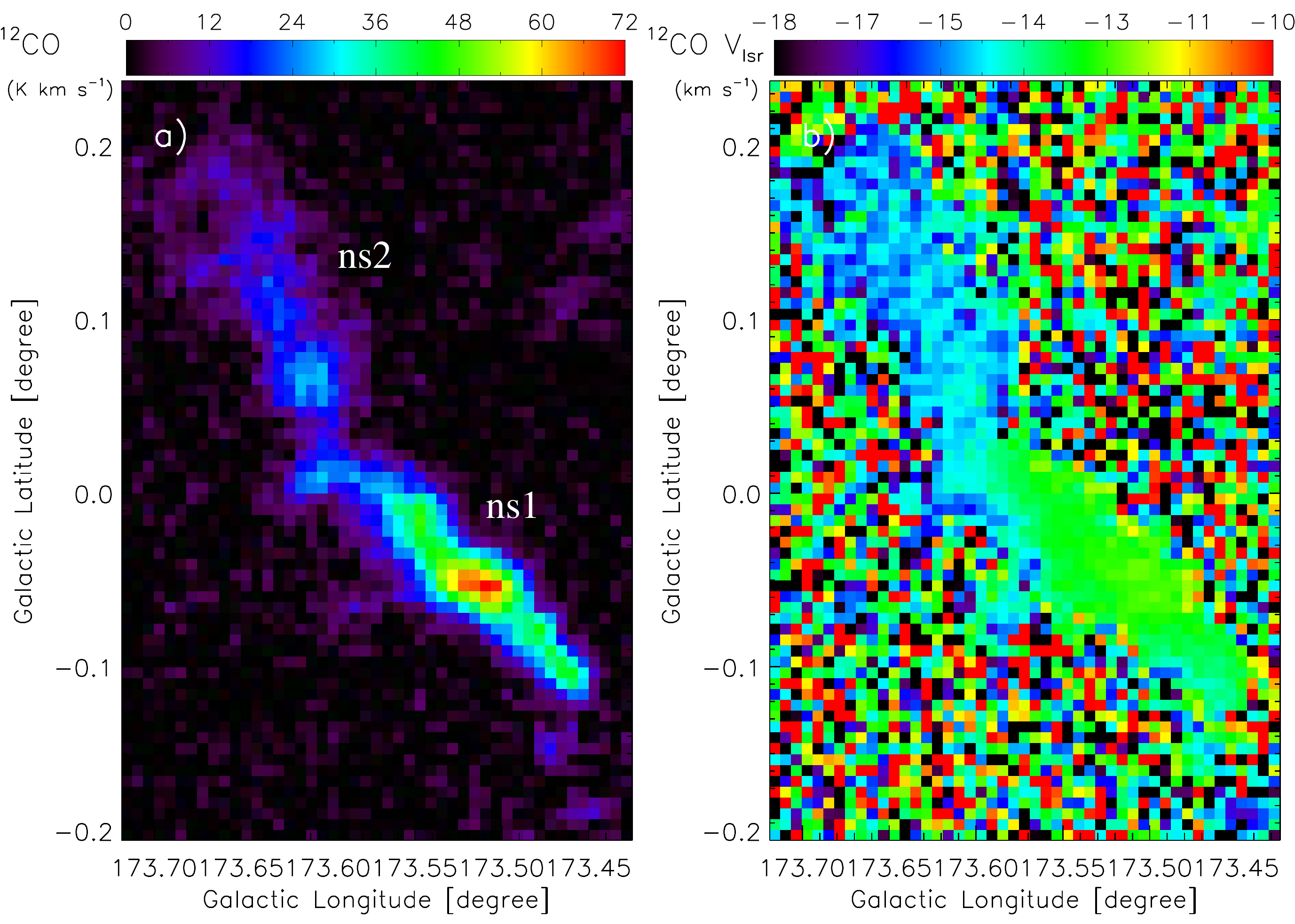}
\epsscale{0.95}
\plotone{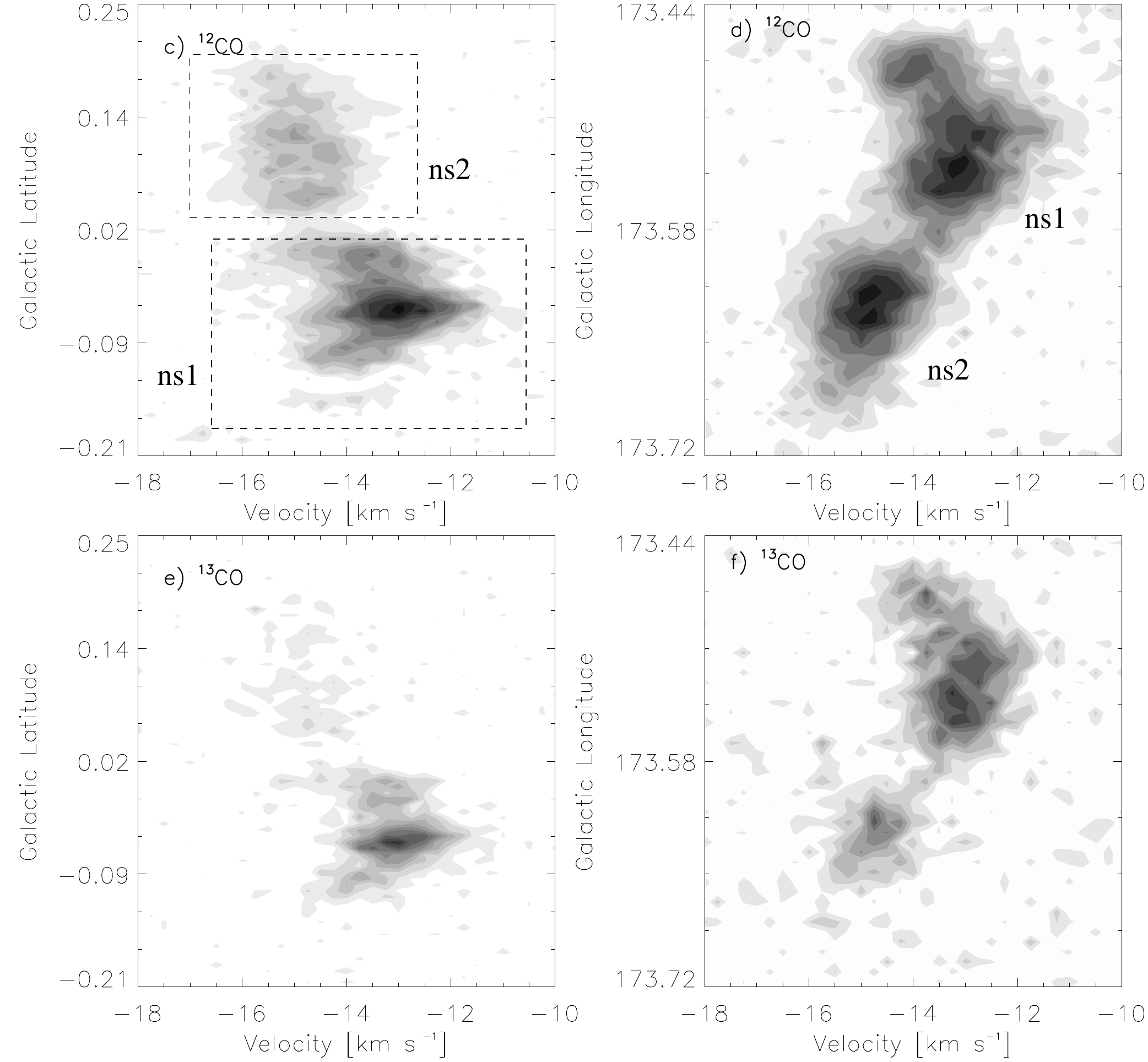}
\caption{a) A zoomed-in integrated $^{12}$CO (J=1-0) emission map toward the elongated features (see a dotted-dashed box in Figure~\ref{sg7}a). 
The $^{12}$CO emission is integrated over a velocity range of $-$17.75 to $-$10 km s$^{-1}$. 
b) The $^{12}$CO first-order moment map. The bar at the top indicates the V$_{lsr}$ (in km s$^{-1}$). 
c) Latitude-velocity plot of $^{12}$CO.
d) Longitude-velocity plot of $^{12}$CO.
e) Latitude-velocity plot of $^{13}$CO.
f) Longitude-velocity plot of $^{13}$CO.
In the panels ``c" and ``e", the CO emission is integrated over 
the longitude from 173$\degr$.44 to 173$\degr$.72.
In the panels ``d" and ``f", the CO emission is integrated over 
the latitude from $-$0$\degr$.21 to 0$\degr$.25.}
\label{sg10}
\end{figure*}

To further study the molecular velocity structures, Figure~\ref{sg8} presents the position-velocity maps in the direction of the site S234.
Figures~\ref{sg8}a and~\ref{sg8}c show the latitude-velocity plots 
of $^{12}$CO and $^{13}$CO, respectively. 
Figures~\ref{sg8}b and~\ref{sg8}d present the longitude-velocity plots of 
$^{12}$CO and $^{13}$CO, respectively. 
The nebulous streams ns1 and ns2 are depicted at different velocity peaks, but they are linked by noticeable intermediate diffuse emission. 
These maps confirm the physical association of different features (i.e. ns2, ns1, dust shell, and dust condensation) in the site S234, and the velocity spread is observed toward all these different features. Using the $^{12}$CO line data, a zoomed-in view of the nebulous streams is displayed in 
Figures~\ref{sg10}a and~\ref{sg10}b. 
The integrated $^{12}$CO emission at [$-$17.75, $-$10] km s$^{-1}$ is shown in 
Figure~\ref{sg10}a, while the first moment map of $^{12}$CO is presented in Figure~\ref{sg10}b. 
The first moment map clearly displays the boundary of the two nebulous streams. 
In the direction of nebulous streams, Figures~\ref{sg10}c --~\ref{sg10}f show the position-velocity 
plots of $^{12}$CO and $^{13}$CO. 
The filaments ns1 and ns2 are interconnected in the velocity space of $^{12}$CO and $^{13}$CO (see broken boxes in Figure~\ref{sg10}a). In the velocity space, a velocity spread is noticeably seen in the direction of ns1. 
\subsubsection{{\it Herschel} column density and temperature maps of S234}
\label{hersec:s234}
\begin{figure*}
\epsscale{0.52}
\plotone{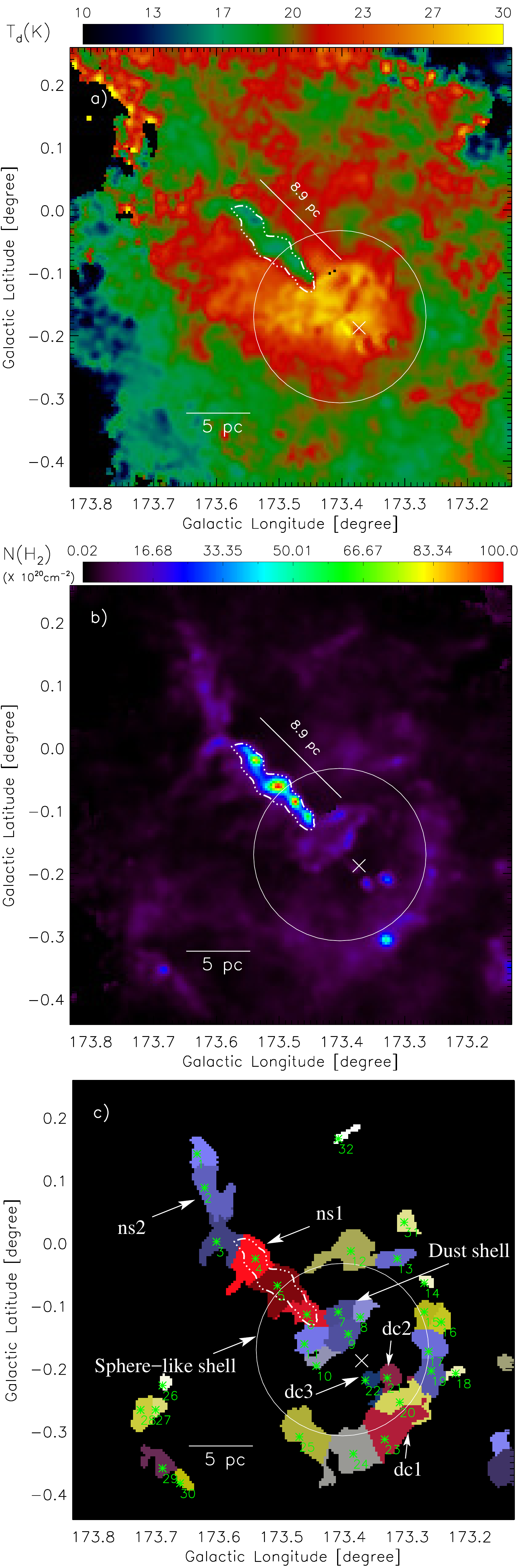}
\caption{a) {\it Herschel} temperature map of the site S234.
b) {\it Herschel} column density map of the site S234. 
c) The positions of the {\it Herschel} clumps are indicated by asterisks. 
The extension of each clump is also highlighted along with its corresponding clump 
ID (see Table~\ref{tab2}).
In each panel, the filament ns1, having length $\sim$8.9 pc, is highlighted by the column density contour with a level of 1.33 $\times$ 10$^{21}$ cm$^{-2}$.} 
\label{sg13}
\end{figure*}

Figures~\ref{sg13}a and~\ref{sg13}b display the temperature and column 
density maps (resolutions $\sim$37$''$) of the site S234. 
In Figure~\ref{sg13}a, the filamentary features are seen with temperatures (T$_{d}$) of around 19 K, 
while Stock 8 and dust shell are depicted in a temperature range 
of 27--30 K (see a big circle in Figure~\ref{sg13}a). 
The column density map reveals the embedded morphology of the S234 complex. 
The map clearly depicts several condensations (having peak $N(\mathrm H_2)$ $\sim$6--10~$\times$~10$^{21}$ 
cm$^{-2}$) in a single $\sim$8.9 pc long filamentary feature (or nebulous stream ns1), 
which is indicated by the column density contour of 1.33 $\times$ 10$^{21}$ cm$^{-2}$. 
The nebulous stream ns2, having length $\sim$8.7 pc, is also traced in the column density map. Interestingly, in the temperature map, the nebulous stream ns1 (around 19 K) appears to be surrounded 
by an extended structure with temperatures of 27--30 K (see Figure~\ref{sg13}a). 
We have identified a total of 32 clumps in the column density map, which are labeled in Figure~\ref{sg13}c. 
Table~\ref{tab2} summarizes the properties (i.e. mass and effective radius) of the identified {\it Herschel} clumps. 
Among these clumps, we find that 25 clumps are distributed toward the sphere-like shell 
structure (including the dust condensations), nebulous streams, 
and dust shell (see ID nos. 1--25 in Table~\ref{tab2}). 
The masses of these clumps vary between 10--593 M$_{\odot}$. 
Figure~\ref{sg13}c also displays the extension of each clump. 
Twelve Clumps (IDs 12-20 and 23--25) are seen toward the sphere-like shell, while the dust shell contains five clumps (IDs 7--11). 
The condensation dc1 (see clump ID 23) is found to be more massive compared to other 
condensations (i.e. dc2 (see clump ID 21) and dc3 (see clump ID 22)). 
Three clumps (IDs 1--3) are associated with the filament ns2 (length $\sim$8.7 pc, temperature $\sim$19 K), 
and the total mass of these clumps is 392 M$_{\odot}$.
The filament ns1 (length $\sim$8.9 pc, temperature $\sim$19 K) contains three clumps (IDs 4--6), 
and the total mass of these clumps is $\sim$1154 M$_{\odot}$, which we also consider as the 
mass of the filament ns1 (i.e. M$_{ns1}$). 
Note that at least four column density peaks (or condensations) are found within these three clumps. 

The values of $M_{\rm line,obs}$ for the filaments ns1 and ns2 are also computed to be 
$\sim$130 and $\sim$45 M$_{\odot}$ pc$^{-1}$, respectively. 
These values are much higher than $M_{\rm line,crit}$ $\sim$32 M$_{\odot}$ pc$^{-1}$ (at T = 20 K). 
Hence, there are two thermally supercritical filaments in the site S234. 
\subsubsection{Embedded young stellar objects in S234}
\begin{figure*}
\epsscale{1.05}
\plotone{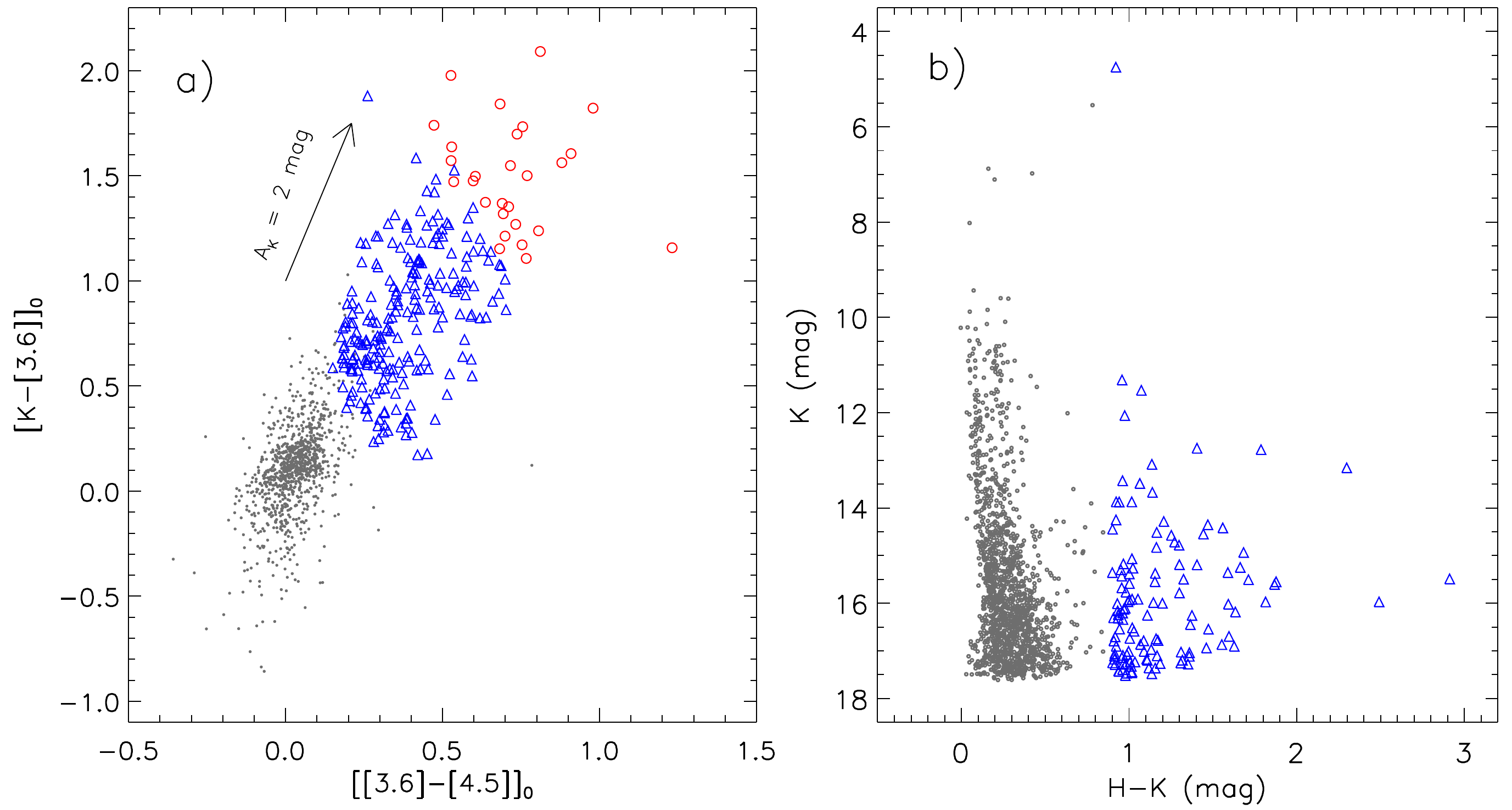}
\caption{a) Dereddened color-color ([K$-$[3.6]]$_{0}$ $vs$ [[3.6]$-$[4.5]]$_{0}$) plot 
of point-like objects toward the selected area around S234 (see Figure~\ref{sg4}a). 
The extinction vector is also drawn in the panel \citep[e.g.][]{flaherty07}. 
b) Color-magnitude (H$-$K/K) plot of point-like objects. 
In each panel, circles and triangles indicate Class~I and Class~II YSOs, respectively. 
In both the panels, the dots (in gray) present the stars with only photospheric emissions. 
We have randomly plotted only some of these stars in both the panels.}
\label{sg14}
\end{figure*}

\begin{figure*}
\epsscale{0.7}
\plotone{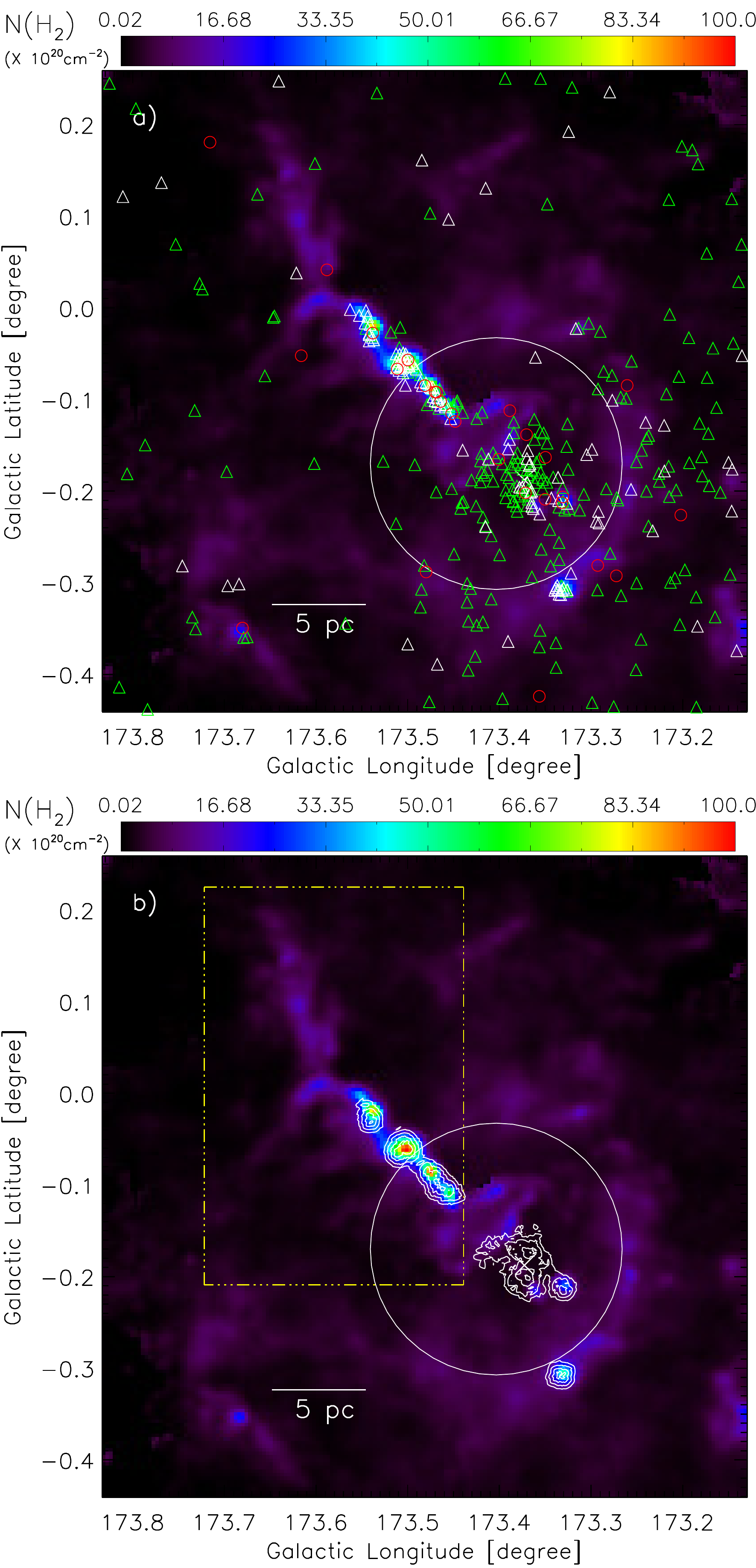}
\caption{a) Overlay of selected YSOs on the {\it Herschel} column density map. 
The YSOs highlighted with red circles and green triangles are selected based on the dereddened color-color scheme (see Figure~\ref{sg14}a), 
while white triangles refer the YSOs identified using the color-magnitude plot (see Figure~\ref{sg14}b). 
b) Overlay of surface density contours of YSOs (in white) on the {\it Herschel} column density map. The contour levels are 3, 5, 10, and 25 YSOs pc$^{-2}$.}
\label{sg15}
\end{figure*}

Figures~\ref{sg14}a and~\ref{sg14}b display the dereddened color-color plot 
(i.e. [K$-$[3.6]]$_{0}$ and [[3.6]$-$[4.5]]$_{0}$) and the NIR color-magnitude plot 
(i.e. H$-$K/K) of point-like objects in the site S234, respectively. 
Color-color and color-magnitude schemes give 251 (27 Class~I and 224 Class~II) YSOs, and 122 YSOs, respectively. Figure~\ref{sg15}a shows the spatial distribution of the YSOs overlaid on the column density map. Figure~\ref{sg15}b presents the overlay of the surface density contours 
of YSOs on the column density map. The YSO density contour levels are 3, 5, 10, and 25 YSOs pc$^{-2}$. 
More detailed descriptions for obtaining the surface density map 
are given in \citet{dewangan17a}. 
The clusters of YSOs are seen toward the filament ns1, Stock 8, and 
dust condensations (i.e. dc1, dc2, and dc3), revealing ongoing star formation activities toward them. 
The condensation dc2 contains both the Class~I and Class~II YSOs, while only Class~II YSOs are 
found toward dc1 and dc3. 
\subsubsection{Embedded condensations in the filament ns1}
\label{hsec:fn1}
\begin{figure*}
\epsscale{1}
\plotone{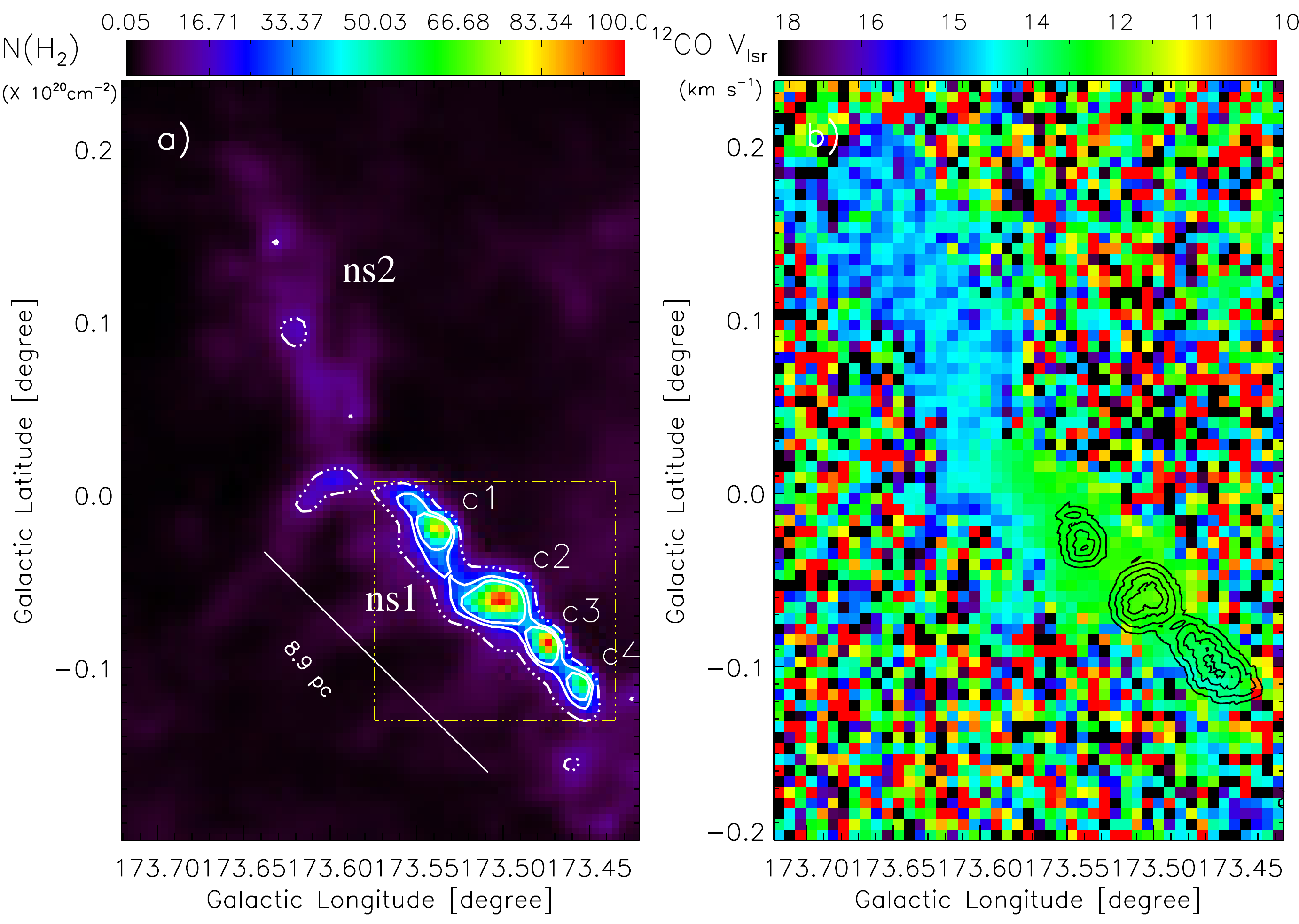}
\epsscale{0.9}
\plotone{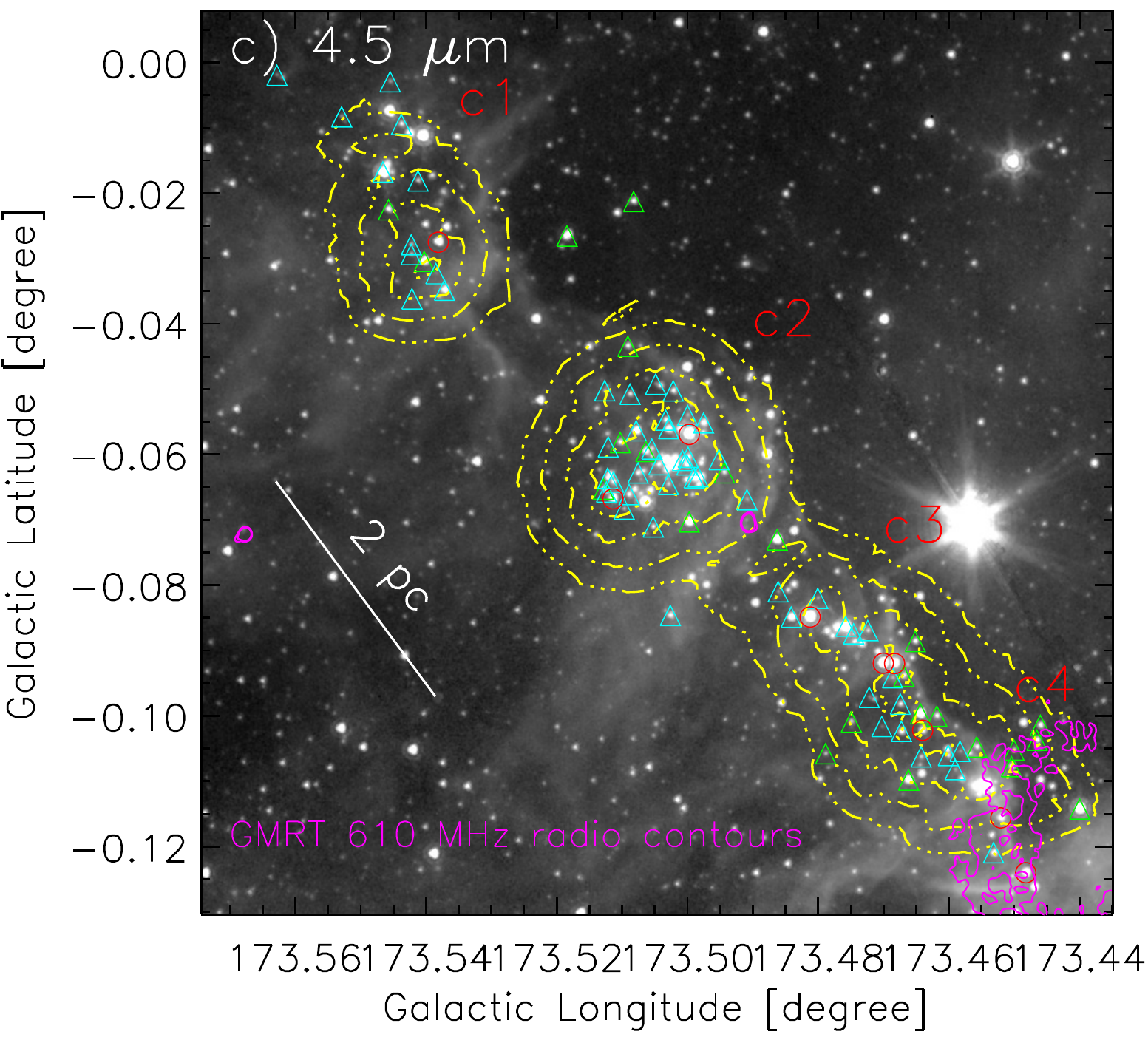}
\caption{a) A zoomed-in view of the {\it Herschel} column density map toward the elongated features (see a dotted-dashed box in Figure~\ref{sg15}b).
The column density contours with the levels of 10$^{20}$ cm$^{-2}$ $\times$ (13.3, 28.3, 38.5) are also drawn in the map. 
The peaks of four condensations, designated as c1-c4, are also highlighted in the map. 
b) Velocity-field (moment 1) map of $^{12}$CO overlaid with the surface density contours of YSOs (in black), which are the same as in Figure~\ref{sg15} (see also Figure~\ref{sg10}b). 
c) Overlay of YSO surface density and GMRT 610 MHz radio continuum emission contours on the {\it Spitzer} 4.5 $\mu$m image (see a dotted-dashed box in Figure~\ref{sg16}a).}
\label{sg16}
\end{figure*}

Figures~\ref{sg16}a and~\ref{sg16}b display the column density 
and first moment maps of the filaments ns1 and ns2. 
The surface density contours of YSOs are also overlaid on the first moment map. 
The supercritical filament ns2 is associated with three clumps (nos. 1--3), where YSOs are not found. 
No radio continuum emission is observed toward the filament ns2. 
It indicates that the filament ns2 seems to be at an early stage of fragmentation. 

On the other hand, the supercritical filament ns1 hosting three clumps (nos. 4--6) is seen with a large number of YSOs. 
In the column density map (resolutions $\sim$37$''$ or 0.5 pc at a distance of 2.8 kpc), 
we find at least four distinct condensations, designated as c1--c4, within three clumps in the direction of the filament ns1. The boundary of each condensation is traced with the column density contour of 
28.3 $\times$ 10$^{20}$ cm$^{-2}$ (see Figure~\ref{sg16}a). The peak column densities of c1, c2, c3, and c4 are found to be $\sim$8.6 $\times$ 10$^{21}$ cm$^{-2}$ ($A_V$ = 9.2 mag), $\sim$10 $\times$ 10$^{21}$ cm$^{-2}$ ($A_V$ = 10.7 mag), $\sim$10 $\times$ 10$^{21}$ cm$^{-2}$ ($A_V$ = 10.7 mag), and $\sim$6 $\times$ 10$^{21}$ cm$^{-2}$ ($A_V$ = 6.4 mag), respectively. 
Here, we use a conversion relation, i.e. $A_V=1.07 \times 10^{-21}~N(\mathrm H_2)$ \citep{bohlin78}. 
The masses (radii) of c1, c2, c3, and c4 are computed to be $\sim$130 $M_\odot$ ($\sim$0.6 pc), $\sim$170 $M_\odot$ ($\sim$0.75 pc), $\sim$300 $M_\odot$ ($\sim$0.9 pc), and $\sim$70 $M_\odot$ ($\sim$0.5 pc), respectively. 

Based on the $^{13}$CO line data, Mach numbers are derived for the condensations c1, c2, c3, and c4 to be 2.4, 3.1, 2.7, and 2.2, respectively, indicating that all the condensations are supersonic. 
Mach number is defined by the ratio of non-thermal velocity dispersion ($\sigma_{\rm NT}$) to sound speed ($a_{s}$). We have also computed the ratio of thermal to 
non-thermal pressure \citep[$P_{TNT} = {a_s^2}/{\sigma^2_{NT}}$;][]{lada03}. 
The values of $P_{TNT}$ for c1, c2, c3, and c4 are estimated to be 0.17, 0.11, 0.14, and 0.20, 
respectively. It suggests that the non-thermal pressure is higher than the thermal pressure in all the condensations. One can estimate the sound speed $a_{s}$ (= $(k T_{kin}/\mu m_{H})^{1/2}$) 
using the value of gas kinetic temperature (T$_{kin}$) and mean molecular weight ($\mu$=2.37; approximately 70\% H and 28\% He by mass).
The non-thermal velocity dispersion is given by:
\begin{equation}
\sigma_{\rm NT} = \sqrt{\frac{\Delta V^2}{8\ln 2}-\frac{k T_{kin}}{29 m_H}} = \sqrt{\frac{\Delta V^2}{8\ln 2}-\sigma_{\rm T}^{2}} ,
\label{sigmanonthermal}
\end{equation}
where $\Delta V$ is the measured Full Width Half Maximum (FWHM) linewidth of the observed $^{13}$CO spectra, 
$\sigma_{\rm T}$ (= $(k T_{kin}/29 m_H)^{1/2}$) is the thermal broadening for $^{13}$CO at T$_{kin}$, and $m_H$ is the mass of hydrogen atom. 
The measured $\Delta V$ values for c1, c2, c3, and c4 
are 1.5, 1.8, 1.6, and 1.4 km s$^{-1}$, respectively. 
In the calculation, we have used the value of T$_{kin}$ to be $\sim$19 K (see Section~\ref{hersec:s234}). The estimated values of $\sigma_{\rm NT}$ for c1, c2, c3, and c4 are 
0.62, 0.78, 0.69, and 0.57 km s$^{-1}$, respectively. 
The ratio of $\sigma_{\rm NT}$/$a_{s}$ (i.e. Mach number) is also calculated for the filament ns1 to be $\sim$3, indicating that the thermally supercritical filament ns1 is supersonic. One can also consider the contribution of the non-thermal gas motions in the estimate of the critical line mass ($M_{\rm line,crit}$), which is known as the 
virial line mass \citep[i.e. $M_{\rm line,vir}$;][]{andre14,kainulainen16}. 
One can write the expression of $M_{\rm line,vir}$ = [1 + ($\sigma_{\rm NT}$/$a_{s}$)$^{2}$] $\times$ [16~M$_{\odot}$ pc$^{-1}$ $\times$ (T$_{gas}$/10 K)]. In the case of the filament ns1, the value of $M_{\rm line,vir}$ ($\sim$320 M$_{\odot}$ pc$^{-1}$) is more than 10 times higher than $M_{\rm line,crit}$ ($\sim$32 M$_{\odot}$ pc$^{-1}$ at T = 20 K), and is also larger than $M_{\rm line,obs}$ ($\sim$130 M$_{\odot}$ pc$^{-1}$). The significance of these estimates is discussed in Section~\ref{sec:discx2}.

Using the {\it Spitzer} 4.5 $\mu$m image, a zoomed-in view of the ns1 is shown in Figure~\ref{sg16}c. 
The image is also overlaid with the YSO surface density contours, 
the positions of 88 YSOs (9 Class~I and 79 Class~II), and the GMRT radio contours. The radio continuum emission is absent toward all the four condensations. Considering the detections of both the Class~I and Class~II YSOs, we use star formation 
timescale to be 1-2 Myrs in the filament ns1 \citep[e.g.][]{jose08,evans09}. 
We have also determined the star formation efficiency (SFE) and the star formation rate 
(SFR) in the filament ns1. In general, the low-mass star-forming regions follow a log-normal Initial Mass Function 
(IMF) with a characteristic mass of 0.5 $M_\odot$ \citep[e.g.][]{chabrier03}.
Hence, we assume a mean mass of 0.5 $M_\odot$ for each YSO in the filament ns1. 
About five Class~II YSOs appear away from the condensations or surface density contours (see Figure~\ref{sg16}c). Hence, we compute a total mass of 80 YSOs in the filament ns1 
(i.e. M$_{ysos}$ = 80 $\times$ 0.5 = 40 $M_\odot$). 
Adopting the values of M$_{ysos}$ and a total mass of the filament ns1 (i.e. M$_{ns1}$ = 1154 $M_\odot$), we have obtained the SFE (i.e. M$_{ysos}$/(M$_{ysos}$ + M$_{ns1}$)) to be $\sim$0.033 (or $\sim$3.3\%). 
Following the work of \citet{evans09}, the SFR (i.e. M$_{ysos}$/t$_{sf}$) is 
estimated to be $\sim$40--20 $M_\odot$ Myr$^{-1}$, where t$_{sf}$ is the star formation timescale (i.e. $\sim$1--2 Myr). This quantitative analysis indicates the intense star formation activity in the thermally supercritical filament ns1. 
\section{Discussion}
\label{sec:disc}
\subsection{Large-scale structure in the direction of {\it l} = 172$\degr$.8; {\it b} = 1$\degr$.5}
\label{sunsec:disc1}
\begin{figure*}
\epsscale{0.83}
\plotone{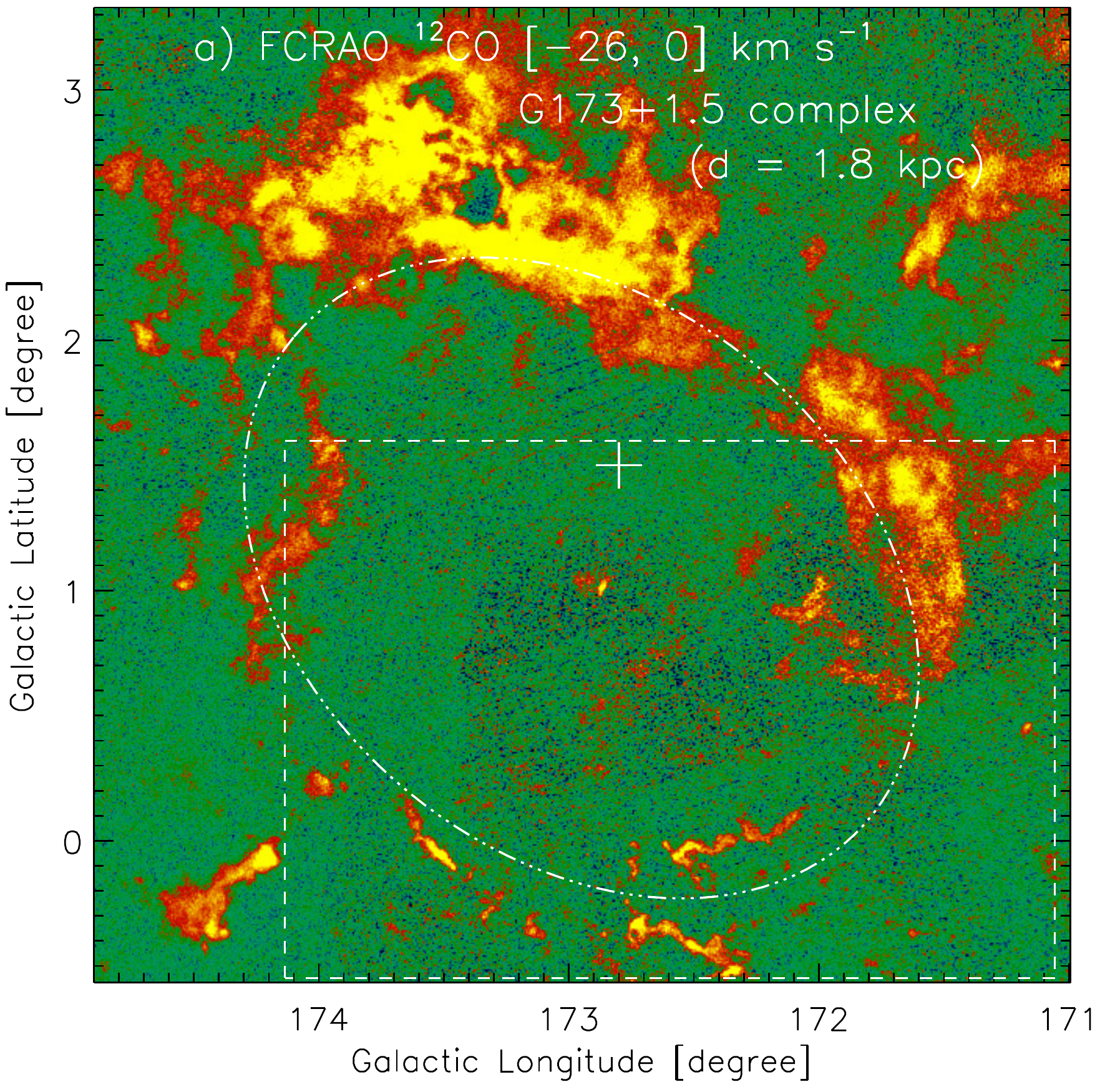}
\epsscale{0.83}
\plotone{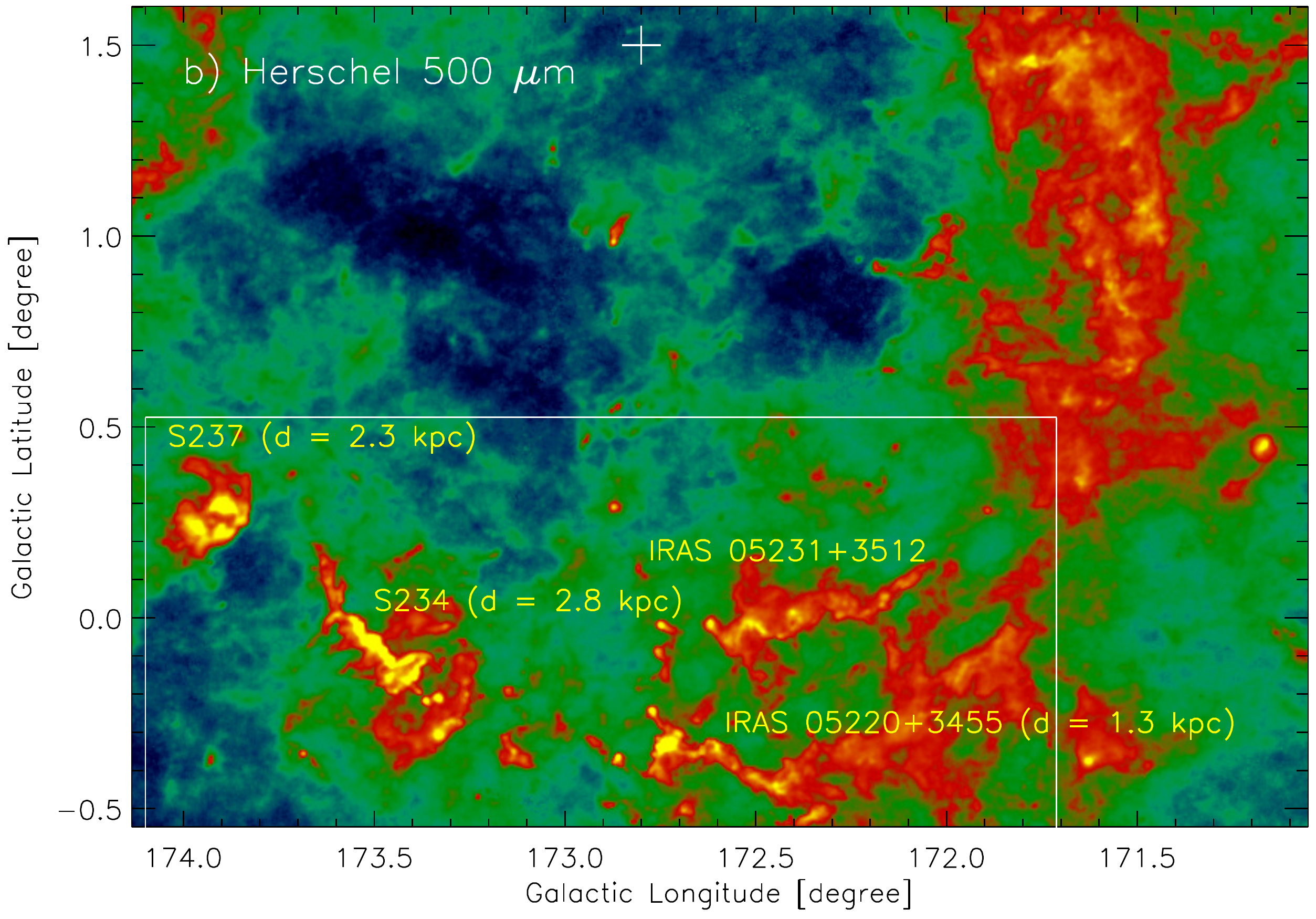}
\caption{a) Large-scale integrated $^{12}$CO (J=1-0) emission map in the direction 
of {\it l} = 171$\degr$.0--174$\degr$.9; {\it b} = $-$0$\degr$.567--3$\degr$.323. 
The position (at {\it l} = 172$\degr$.8; {\it b} = +1$\degr$.5) is highlighted by a plus. 
The $^{12}$CO emission is integrated over a velocity range of $-$26 to 0 km s$^{-1}$. 
A dotted-dashed ellipse indicates an extended shell-like structure/bubble-shaped nebula in 
the direction of {\it l} = 172$\degr$.8; {\it b} = +1$\degr$.5 \citep[e.g.][]{kang12,kirsanova17}. 
An area shown in Figure~\ref{ng16}b is highlighted by a dashed box. 
In the Galactic nothern direction, a complex G173+1.5 (at d = 1.8 kpc) is also labeled. 
b) {\it Herschel} 500 $\mu$m image in the direction 
of {\it l} = 171$\degr$.0--174$\degr$.1; {\it b} = $-$0$\degr$.548--1$\degr$.6. 
Different sites (i.e. S237, S234, IRAS 05231+3512, and IRAS 05220+3455) along with their distances are 
labeled in the figure (see also Figure~\ref{sg1}a). 
A solid box shows an area which is presented in Figure~\ref{sg1}a. The plus symbol is the same as in Figure~\ref{ng16}a.} 
\label{ng16}
\end{figure*}

In Figure~\ref{ng16}a, we present an integrated $^{12}$CO emission map toward {\it l} = 172$\degr$.8; {\it b} = 1$\degr$.5, and the molecular emission is integrated over a velocity interval of [$-$26, 0] km s$^{-1}$. 
An extended structure is seen in the molecular intensity map (see a dotted-dashed ellipse in Figure~\ref{ng16}a). 
Earlier, \citet{kang12} and \citet{kirsanova17} also reported a similar extended shell-like structure/bubble-shaped nebula in the same longitude direction \citep[see Figure~5 in][and Figure~1 in Kirsanova et al. 2017]{kang12}. In Figure~\ref{ng16}b, the Galactic southern side of the extended structure is also presented using the {\it Herschel} 500 $\mu$m image, where our selected star-forming regions (i.e. S234 (V$_{lsr}$ $\sim$$-$13.4 km s$^{-1}$; d = 2.8 kpc), IRAS 05220+3455 (V$_{lsr}$ $\sim$$-$10 km s$^{-1}$; d = 1.3 kpc), IRAS 05231+3512 (V$_{lsr}$ $\sim$$-$6 km s$^{-1}$), and S237 (V$_{lsr}$ $\sim$$-$4.5 km s$^{-1}$; d = 2.3 kpc)) are located (see a solid box in Figure~\ref{ng16}b). 
An active star-forming complex G173+1.5 is located in the Galactic northern side of the extended structure (at d $\sim$1.8 kpc and 
V$_{lsr}$ $\sim$ $-$20 km s$^{-1}$), where five Sharpless H\,{\sc ii} regions S231--S235 are distributed. 
These star-forming regions appear to be distributed toward the edges of the extended molecular structure. 
However, in the direction of the extended structure, different velocity components/star-forming sites are traced in the velocity range of [$-$26, 0] km s$^{-1}$, and are situated at different distances. 
It implies that the extended molecular structure cannot be considered as a single physical system. 

Previously, \citet{kang12} reported the existence of an old supernova remnant (SNR) FVW 172.8+1.5 
(age $\sim$0.3 Myr) in the direction of {\it l} = 172$\degr$.8; {\it b} = 1$\degr$.5, and 
suggested that the SNR is responsible for the extended shell-like structure/bubble-shaped observed in the same longitude direction (see Figure~5 in their paper). They adopted a systemic velocity (V$_{sys}$) and a distance of FVW172.8+1.5 to be $\sim$$-$20 km s$^{-1}$ 
and $\sim$1.8 kpc, respectively, which are very similar to that of the H\,{\sc ii} complex G173+1.5. 
Hence, these authors pointed out that the SNR is originated inside the H\,{\sc ii} complex G173+1.5. 
They proposed that the energetics of the SNR (having a kinetic energy $\sim$2.5 $\times$ 
10$^{50}$ erg) may have affected the star formation activities in the H\,{\sc ii} complex G173+1.5 and our selected star-forming sites. However, the SNR with the age of $\sim$0.3 Myr is too young to be a triggering agent. 
It is supported with the fact that the average ages of YSOs are computed to be $\sim$0.44--3 Myr \citep[e.g.][]{evans09}, and YSOs are utilized as an excellent tracer for inferring the 
star formation activity in a given star-forming site. 
\citet{kang12} further proposed that the parent stellar association of the SNR could have influenced the birth of the massive OB stars seen near the edges of the extended structure. 
However, \citet{kirsanova14} discussed that shock waves from the SNR cannot trigger the formation of S235. 
Based on the photometric and spectroscopic study, \citet{marco16} also rejected any influence of the large-scale shocks to the star-formation in the site S234. 
Considering the observed distances and molecular velocity ranges, we do not find any physical association of our selected sites with 
the SNR or its parent stellar association \citep[see also][]{marco16}. 
Together, the proposed scenario of \citet{kang12} is unlikely to be applicable in our selected star-forming sites 
(i.e. S234, V582, IRAS 05220+3455, and S237).
\subsection{Supercritical filaments in the direction of {\it l} = 171$\degr$.7 -- 174$\degr$.1}
\label{ssssec:discx1}
In the direction of our selected longitude range, the sites S234, V582, IRAS 05220+3455, 
and IRAS 05231+3512 are spatially seen in the infrared images. 
Based on the existing distance estimates \citep[e.g.][]{marco16} and our molecular line data analysis, 
we find that these sites are not part of the same physical system (see also Section~\ref{sunsec:disc1}).  
Using the {\it Herschel} sub-mm continuum images, we have identified 
at least five filamentary features (ff1, ff2, ff3, ns1, and ns2) toward {\it l} = 171$\degr$.7 -- 174$\degr$.1.
The site IRAS 05231+3512 is found to be embedded in the filament ff1, while the filament ff2 contains the sites V582 and IRAS 05220+3455. In the site S234, two embedded filaments (i.e. ns1 and ns2) are investigated in the direction of previously known nebulous stream feature. 
One can note that in the literature, there is no estimate of the distance of the site IRAS 05231+3512. However, in this work, 
we have used the distance of 1.3 kpc to the IRAS site. As highlighted earlier, the sites IRAS 05231+3512 
(V$_{lsr}$ $\sim$$-$6 km s$^{-1}$) and IRAS 05220+3455 (V$_{lsr}$ $\sim$$-$10 km s$^{-1}$; d = 1.3 kpc) do not belong to a 
single system (see Section~\ref{subsec:eff}). 
Hence, it is possible that the site IRAS 05231+3512 could be located at a distance ranging from 1.3 to 2.0 kpc. 
This distance range can supported with the fact that there is another nearby open cluster NGC 1907 ({\it l} = 172$\degr$.6; {\it b} = +0$\degr$.3; V$_{lsr}$ $\sim$0.1) located at a distance of $\sim$1.8 kpc \citep[e.g.,][]{marco16}. 

In order to study the filaments, the observed line masses of filaments are computed, and are compared with a critical line mass in an equilibrium state. 
As discussed in Section~\ref{subsec:v582}, the line mass of an infinite and isolated cylinder in equilibrium between thermal and gravitational pressures can be determined using a given 
temperature value \citep[e.g.][]{ostriker64}. The thermally supercritical filaments follow a condition $M_{\rm line,obs}$ $>$ $M_{\rm line,crit}$, whereas thermally subcritical filaments agree with a condition $M_{\rm line,obs}$ $<$ $M_{\rm line,crit}$. 
The supercritical filaments can collapse radially perpendicular to the main long axis of the filaments. 
In this context, gravitational fragmentation is one of the important physical processes 
to explain the observed clumps in filaments \citep[e.g.][]{ostriker64,inutsuka97,hartmann02,arzoumanian13,andre14,kainulainen16,williams18}. 
With the help of the {\it Herschel} column density and temperature maps, the features ff1, ff2, ff3, ns1, and ns2 are consistent with the condition of thermally supercritical filaments. 
In recent years, researchers are also trying to explore embedded filaments at an early stage of fragmentation, where star formation activities are not yet started or there is very little sign of ongoing star formation \citep[e.g.][]{kainulainen16}. Such filaments may preserve the initial conditions of gravitational fragmentation. A comparison study of YSOs against clumps in filaments can provide an important clue to assess different evolutionary stages of filaments. More detailed discussion concerning star formation in the thermally supercritical filaments is presented in Sections~\ref{sec:discx1} and~\ref{sec:discx2}.  
\subsection{Star formation scenario in the sites V582, IRAS 05220+3455, and IRAS 05231+3512}
\label{sec:discx1}
In the $\sim$1$\degr$.17 $\times$ 0$\degr$.81 area of V582, three filaments ff1, ff2, and ff3 are identified, and are classified as thermally supercritical. 
The column density map has shown the existence of condensations/clumps in these filaments.
Hence, radial collapse scenario appears to be operated in these filaments. 
We have found less number of infrared-excess sources in the two filaments ff1 and ff3, where the radio continuum emission is also not detected. It implies that the filament ff1 harboring IRAS 05231+3512 may be very young, and seems to be located toward the Galactic southern edge of the extended molecular structure (see Figures~\ref{ng16}a and~\ref{ng16}b). Hence, considering the distance 
uncertainty of this IRAS site (see Section~\ref{ssssec:discx1}) and the age of the young SNR (see Section~\ref{sunsec:disc1}), 
the formation of the filament ff1 could be influenced by the SNR shock. 
In Figure~\ref{vsg4}b, the observed velocity gradients along the filament ff1 appear to indicate the accretion in the filamentary 
cloud \citep[e.g.][]{vazquez09,arzoumanian13}, however protostellar feedback cannot be completely ignored for the velocity gradient in the filament ff1. 
In the thermally supercritical filament ff2, a group of YSOs is seen around IRAS 05220+3455, and a cluster of YSOs and at least five radio continuum peaks are also observed toward the site V582. 
The associated cometary globules and bright-rimmed clouds with the site V582 are seen at one end of ff2. 
Recently, \citet{kun17} suggested that star formation in the nearby environment of V582 might be 
triggered by the radiation field of a few hot members of Aur OB1 (i.e. HD 35633 or LS V +34$\degr$23). 
However, \citet{marco16} found the object HD 35633 as the main powering source for the site S234. 
Hence, star formation in the site V582 cannot be affected by the object HD 35633. Considering the feedback of massive stars, the presence of massive B type stars may be responsible for the bright-rim clouds and cometary globules in the site V582. 

Based on the number of detected YSOs in the three filaments (see Section~\ref{subsec:v582}), we suggest that the filaments ff1--ff3 and ff2 represent two different evolutionary stages. 
The filament ff2 appears more evolved compared to other two filaments. Taking into account the existence of the supercritical filaments, our results favour the onset of radial collapse process in the filaments ff1, ff2, and ff3.
\subsection{Star formation scenario in the site S234}
\label{sec:discx2}
\begin{figure*}
\epsscale{1}
\plotone{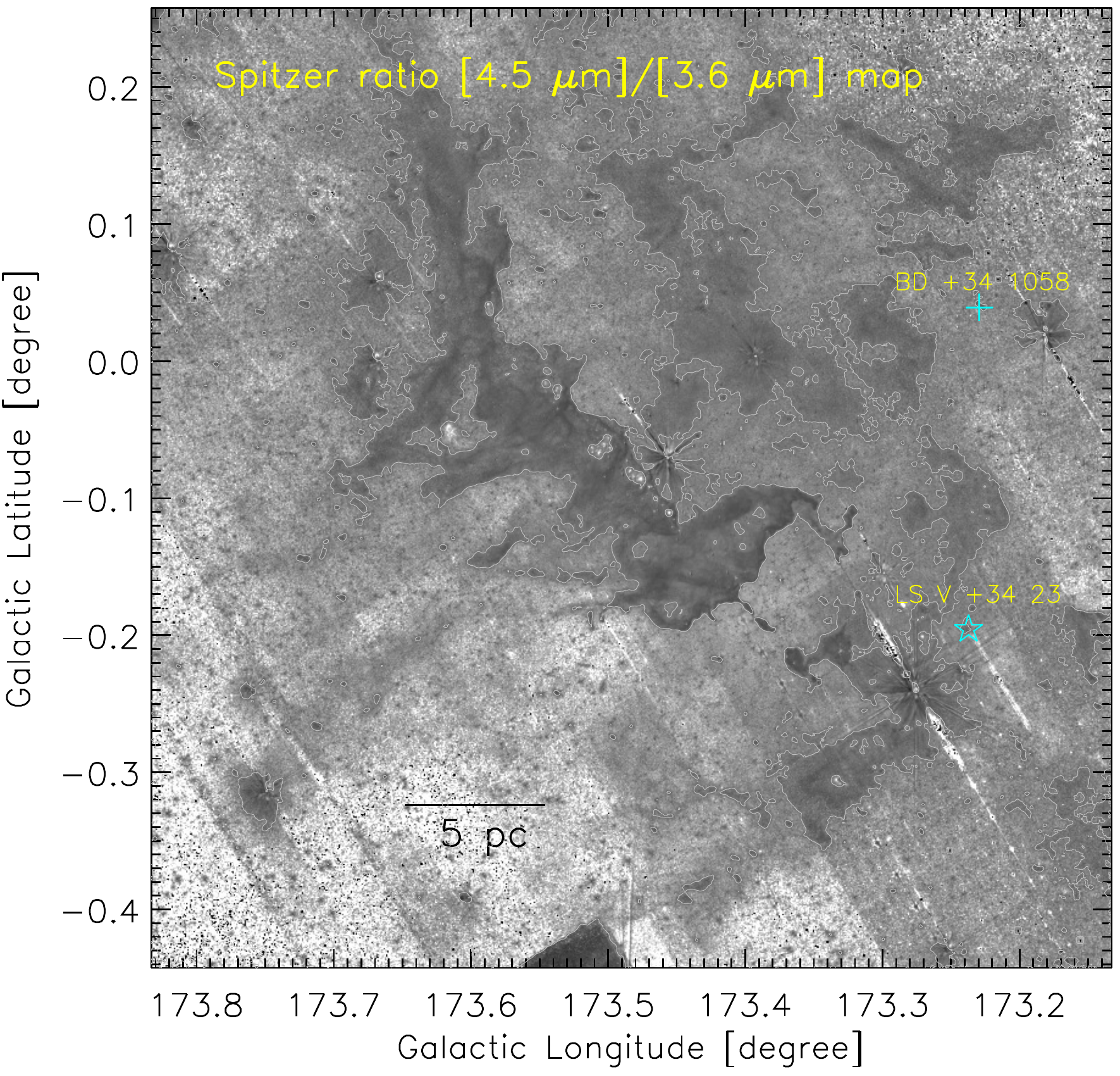}
\caption{The {\it Spitzer} ratio map of 4.5 $\mu$m/3.6 $\mu$m emission toward the site S234. 
The contour of the ratio map of 4.5 $\mu$m/3.6 $\mu$m emission is also shown with a level of 0.85. 
The ratio map is exposed to a Gaussian smoothing function with a width of 4 pixels. 
The positions of stars HD 35633/LS V +34$\degr$23 and BD +34$\degr$1058 are highlighted by star and cross symbols, respectively.}
\label{sg17}
\end{figure*}

In Section~\ref{subsec:s234}, we studied the molecular cloud associated with the site S234 in velocities of [$-$17.75, $-$10] km s$^{-1}$, which hosts the nebulous stream, dust shell, young open cluster Stock 8, and dust condensations. Several OB stars were reported in the cluster Stock 8 by \citet{marco16}. 
\citet{marco16} found the object LS V +34$\degr$23/HD 35633 as the main exciting source for S234 (see the position of this object in Figure~\ref{sg17}). 
The GMRT radio continuum emission at 610 MHz is detected toward the dust shell, Stock 8, 
and dust condensations. 

Earlier works suggested an age of 4--6 Myr for the open cluster Stock 8 \citep[e.g.][]{marco16}, 
and the lifetime of star formation in the nebulous stream was reported to be $\sim$1 Myr \citep[e.g.][]{jose08}. 
\citet{jose08} argued that stars in Stock 8 represent second generation star formation activity, which was 
triggered by the first-generation stars in the site S234. 
The nebulous stream was classified as an active region of recent star formation, where the small obscured cluster CC 14 and a significant number of YSOs were identified \citep[e.g.][]{jose08,marco16}.
It was suggested that the ionizing feedback of OB stars in the Stock 8 (and 
the first-generation stars) has not reached the nebulous stream. Hence, these OB 
stars may not influence the star formation in the nebulous stream \citep[e.g.][]{jose08}.
The nebulous stream was characterized as a bright rim or an ionization front, and was excited by 
a massive O type star (i.e. BD +34$\degr$1058) in the site S234 \citep[e.g.][]{jose08}. They suggested that this source 
was not part of the Stock 8 and the first-generation stars (see the position of this source in Figure~\ref{sg17}).

In this work, we have identified two thermally supercritical filaments ns1 and ns2 toward the nebulous stream (see Section~\ref{hersec:s234}). 
The $\sim$8.9 pc long filament ns1 (at T$_{d}$ $\sim$19 K) is identified as the most interesting feature in the site S234, and is surrounded by an extended temperature structure at T$_{d}$ = 27-30 K. 
The observed temperature structure can be explained by the 
heating from the massive OB stars in the site S234. 
We find that molecular gas in the filament ns1 (around $-$13 km s$^{-1}$) is red-shifted with respect to the molecular gas in other features in the site S234 (around $-$15 km s$^{-1}$). 
It implies that the filament ns1 appears away from the other features in the site S234. 
Additionally, the filament ns1 is relatively more denser region in the molecular cloud associated with the site S234. 
Because of these reasons, the pre-existing molecular gas in the filament ns1 is not destroyed by the feedback of 
massive OB stars in the site S234. But at the same, the filament also seems to be influenced by these massive stars 
(see Figure~\ref{sg17}). In Figure~\ref{sg17}, we present the {\it Spitzer}-IRAC ratio map of 4.5 $\mu$m/3.6 $\mu$m emission in 
the direction of the site S234. As discussed in Section~\ref{subsec:v582}, the dark or gray regions in the ratio map help us to trace the PDRs in the site S234, suggesting the impact of massive stars in their surroundings. 

The supercritical filament ns1 contains three clumps, which further hosts at least four condensations.
These condensations are distributed along the filament. 
The intense star formation activity (having SFE $\sim$3.3\% and SFR $\sim$40--20 M$_\odot$ Myr$^{-1}$) is traced in the filament ns1. Furthermore, the thermally supercritical filament ns1 is also supersonic. 
In Section~\ref{hsec:fn1}, we find $M_{\rm line,vir}$ $>$ $M_{\rm line,obs}$ for the filament ns1. 
It indicates that the feedback of massive stars as an external agent may provide the necessary support to the filament ns1 against the turbulence, and may be responsible for the intense star formation activities in the filament. 
It is also favoured by the presence of the observed age gradient in the site S234. 
More recently, \citet{xu18} also found similar results in the filamentary molecular cloud G47.06+0.26. 
Our interpretation is in agreement with the radiation-driven implosion \citep[RDI; see][]{bertoldi89,lefloch94} process, 
where the expanding ionized region initiates the instability and helps for collapsing the 
pre-existing dense regions in a given molecular cloud.

On the other hand, no star formation is seen in the thermally supercritical filament ns2 containing three {\it Herschel} clumps. Hence, the filament ns2 seems to be at an early stage of fragmentation. 
To obtain more insights into the physical conditions of the filaments, our multi-wavelength data analysis demands the high density tracer and high resolution molecular line observations for the site S234. 
\section{Summary and Conclusions}
\label{sec:conc}
Using a multi-wavelength approach, we present an extensive study of star-forming sites, 
IRAS 05231+3512, IRAS 05220+3455, V582, and S234 located toward {\it l} = 171$\degr$.7 -- 174$\degr$.1; {\it b} = $-$0$\degr$.6 -- 0$\degr$.52. The goal of the present work is to understand the ongoing physical processes in these sites. The major results of this paper are:\\
$\bullet$ The molecular clouds associated with the sites S234, IRAS 05231+3512, and V582 are traced in the velocity ranges of [$-$17.75, $-$10] km s$^{-1}$, [$-$8.5, $-$1] km s$^{-1}$, 
and [$-$12.5, $-$7.5] km s$^{-1}$, respectively. 
Based on the observed radial velocities and available distance estimates, 
we find that these three sites do not belong to a single physical system.\\ 
$\bullet$ Based on the {\it Herschel} data of the $\sim$1$\degr$.17 $\times$ 0$\degr$.81 
area of V582, at least three filamentary features (i.e. ff1 (length $\sim$9.5 pc; mass $\sim$355 M$_{\odot}$), ff2 (length $\sim$12.7 pc; mass $\sim$362 M$_{\odot}$), and ff3 (length $\sim$6.9 pc; mass $\sim$182 M$_{\odot}$)) are identified with temperatures of $\sim$14--18 K.\\
$\bullet$ The feature ff1 hosts IRAS 05231+3512, while other two sites IRAS 05220+3455 and V582 are 
embedded in the feature ff2. 
With the help of the CO line data, the sites IRAS 05220+3455 and V582 are found to be physically linked, 
and are not associated with the site IRAS 05231+3512.\\ 
$\bullet$ The observed line masses of the filaments ff1, ff2, and ff3 are derived to be $\sim$37, $\sim$28, and $\sim$26 M$_{\odot}$ pc$^{-1}$, respectively, which exceed the expected critical line mass at T = 10--15 K (i.e. M$_{\rm line,crit}$ = 16--24 M$_{\odot}$ pc$^{-1}$). Hence, the features ff1, ff2, and ff3 can be characterized as thermally supercritical filaments.\\
$\bullet$ The NVSS radio continuum emission at 1.4 GHz is not observed toward 
the filaments ff1 and ff3, where a very little star formation activity is found.\\
$\bullet$ In the filament ff2, a group of YSOs and massive B type stars are depicted toward the site V582, and a cluster of YSOs is also seen around IRAS 05220+3455.\\ 
$\bullet$ The cometary globules and bright-rimmed clouds linked with the site V582 are found 
at one corner of the filament ff2, and can be explained by the feedback of massive B type stars.\\ 
$\bullet$ In the site S234, using the molecular line data, the features ns1, ns2, dust shell, and dust condensation are physically connected in the velocity range of [$-$17.75, $-$10] km s$^{-1}$. However, the molecular gas in the filament ns1 is red-shifted (peaking around $-$13 km s$^{-1}$) with respect to other molecular features (peaking around $-$15 km s$^{-1}$) in the site.\\
$\bullet$ In the site S234, two features (i.e. ns1 (length $\sim$8.9 pc; mass $\sim$1154 M$_{\odot}$; dust temperature $\sim$19 K; $M_{\rm line,obs}$ $\sim$130 M$_{\odot}$ pc$^{-1}$) and ns2 (length $\sim$8.7 pc; mass $\sim$392 M$_{\odot}$; dust temperature $\sim$19 K; $M_{\rm line,obs}$ $\sim$45 M$_{\odot}$ pc$^{-1}$)) are investigated as thermally supercritical filaments, and host several dust clumps.\\ 
$\bullet$ The {\it Herschel} temperature map reveals the filament ns1 around T$_{d}$ $\sim$19 K surrounded by an extended temperature structure at 27-30~K. The ratio of $\sigma_{\rm NT}$/$a_{s}$ (i.e. Mach number) for the filament ns1 is computed to be $\sim$3, indicating that the thermally supercritical filament ns1 is supersonic. Taking into account the turbulence, our analysis reveals $M_{\rm line,vir}$ ($\sim$320 M$_{\odot}$ pc$^{-1}$) $>$ $M_{\rm line,obs}$ ($\sim$130 M$_{\odot}$ pc$^{-1}$) for the filament ns1.\\ 
$\bullet$ At least four condensations (M$_{clump}$ $\sim$70--300 M$_{\odot}$) are observed in the filament ns1, and are not associated with the GMRT 610 MHz continuum emission.\\ 
$\bullet$ The filament ns2 displays almost no star formation activity, indicating at an early stage of fragmentation.\\ 
$\bullet$ An intense star formation activity is observed in the filament ns1, where both the Class~I and Class~II YSOs are identified. 
In the filament ns1, the star formation efficiency and the star formation rate are estimated to be $\sim$3.3\% and $\sim$40--20 M$_{\odot}$ Myr$^{-1}$ (for 1--2 Myr timescale), respectively.\\

Considering the existence of the thermally supercritical filaments in our selected 
longitude range {\it l} = 171$\degr$.7 -- 174$\degr$.1, our results suggest the onset of radial collapse 
process in the filaments ns2, ff1, ff2, and ff3. At least three filaments ff1, ff3, and ns2 appear to be at an early stage of fragmentation, while the filaments ff2 and ns1 are associated with the ongoing star formation activities. 
In the site S234, our observational outcomes favour the triggered star formation in the filament ns1. 
Together, different evolutionary stages of the filaments are investigated toward {\it l} = 171$\degr$.7 -- 174$\degr$.1. 
\acknowledgments 
We thank the anonymous reviewer for constructive comments and suggestions. 
The research work at Physical Research Laboratory is funded by the Department of Space, Government of India.
This work is based on data obtained as part of the UKIRT Infrared Deep Sky Survey. This publication 
made use of data products from the Two Micron All Sky Survey (a joint project of the University of Massachusetts and 
the Infrared Processing and Analysis Center / California Institute of Technology, funded by NASA and NSF), archival 
data obtained with the {\it Spitzer} Space Telescope (operated by the Jet Propulsion Laboratory, California Institute 
of Technology under a contract with NASA). 
The Canadian Galactic Plane Survey (CGPS) is a Canadian project with international partners. 
The Dominion Radio Astrophysical Observatory is operated as a national facility by the 
National Research Council of Canada. The Five College Radio Astronomy Observatory 
CO Survey of the Outer Galaxy was supported by NSF grant AST 94-20159. The CGPS is 
supported by a grant from the Natural Sciences and Engineering Research Council of Canada (NSERC). 
This paper makes use of data obtained as part of the INT Photometric H$\alpha$ Survey 
of the Northern Galactic Plane (IPHAS, www.iphas.org) carried out at the Isaac Newton Telescope (INT).  
The INT is operated on the island of La Palma by the Isaac Newton Group in the Spanish Observatorio del 
Roque de los Muchachos of the Instituto de Astrofisica de Canarias. 
All IPHAS data are processed by the Cambridge Astronomical Survey Unit, at the Institute of 
Astronomy in Cambridge. 
TB acknowledges funding from the National Natural Science Foundation of China through grant 11633005. 
IZ is supported by the Russian Foundation for Basic Research (RFBR) grants No. 17-52-45020 and 18-02-00660. 
\newpage
%
%
\begin{deluxetable}{ccccccc}
\tablewidth{0pt} 
\tablecaption{Ionized clumps detected in the GMRT 610 MHz continuum map toward the site S234 (see Figure~\ref{sg5}d). 
Table lists ionized clump ID, position ({\it l}, {\it b}), deconvolved effective radius of the ionized clump ($R_\mathrm{HII}$), total flux (S${_\nu}$), 
Lyman continuum photons (log$N_\mathrm{uv}$), and radio spectral type. \label{tab1}} 
\tablehead{ \colhead{ID} & \colhead{{\it l}}   & \colhead{{\it b}}  & \colhead{$R_\mathrm{HII}$}& \colhead{S${_\nu}$}& \colhead{log$N_\mathrm{uv}$} & \colhead{Spectral Type}\\
             \colhead{}  &  \colhead{[degree]} & \colhead{[degree]} & \colhead{(pc)}            &\colhead{(in Jy)}   & \colhead{(s$^{-1}$)}         & \colhead{}}
\startdata 

   1     &  	 173.389    & $-$0.155	    &	   5.30   &	1.613	&     47.96	    &	  B0V-O9.5V      \\
   2     &  	 173.339    & $-$0.209	   &	   0.65   &	0.129	&     46.86	    &	   B0.5V-B0V     \\
   3     &  	 173.440    & $-$0.192	   &	   0.35   &	0.035	&     46.30	    &	  B1V-B0.5V      \\
   4     &  	 173.451    & $-$0.126	   &	   0.30   &	0.026	&     46.15	    &	  B1V-B0.5V      \\
   5     &  	 173.385    & $-$0.030	   &	   0.20   &	0.056	&     46.50	    &	   B0.5V        \\

 \enddata  
\end{deluxetable}
\begin{deluxetable}{ccccc}
\tablewidth{0pt} 
\tablecaption{{\it Herschel} clumps detected toward the site S234 and their physical properties (see Figures~\ref{sg13}b and~\ref{sg13}c). 
Table lists {\it Herschel} clump ID, position ({\it l}, {\it b}), deconvolved effective radius of the clump (R$_{c}$), and clump mass (M$_{clump}$). \label{tab2}} 
\tablehead{ \colhead{ID} & \colhead{{\it l}} & \colhead{{\it b}} & \colhead{R$_{c}$}& \colhead{M$_{clump}$}\\
\colhead{} &  \colhead{[degree]} & \colhead{[degree]} & \colhead{(pc)} &\colhead{($M_\odot$)}}
\startdata 
   1 	 &  173.635   &      0.142   &        1.1   &        60      \\
   2 	 &  173.623   &      0.088   &        1.6   & 	    178      \\
   3 	 &  173.604   &      0.002   &        1.5   & 	    154      \\
   4 	 &  173.542   &     -0.025   &        1.7   & 	    365     \\
   5 	 &  173.507   &     -0.068   &        1.8   & 	    593     \\
   6 	 &  173.460   &     -0.114   &        1.3   & 	    196     \\
   7 	 &  173.410   &     -0.110   &        1.0   & 	     85     \\
   8 	 &  173.375   &     -0.118   &        0.9   & 	     42      \\
   9 	 &  173.394   &     -0.145   &        1.3   & 	    119      \\
  10 	 &  173.445   &     -0.196   &        0.9   & 	     39      \\
  11 	 &  173.464   &     -0.161   &        1.2   & 	     83      \\
  12 	 &  173.390   &     -0.013   &        1.8   & 	    154     \\
  13 	 &  173.316   &     -0.025   &        1.1   & 	     63      \\
  14 	 &  173.273   &     -0.064   &        0.5   & 	     10      \\
  15 	 &  173.273   &     -0.110   &        1.0   & 	     48     \\
  16 	 &  173.246   &     -0.126   &        0.9   & 	     38      \\
  17 	 &  173.266   &     -0.173   &        1.2   & 	     85      \\
  18 	 &  173.223   &     -0.208   &        0.5   & 	     13      \\
  19 	 &  173.262   &     -0.204   &        1.3   & 	    100      \\
  20 	 &  173.312   &     -0.254   &        1.5   & 	    100      \\
  21 	 &  173.332   &     -0.215   &        1.0   & 	     72      \\
  22 	 &  173.367   &     -0.219   &        0.8   & 	     40     \\
  23 	 &  173.336   &     -0.313   &        1.9   & 	    323      \\
  24 	 &  173.386   &     -0.336   &        1.9   & 	    185      \\
  25 	 &  173.472   &     -0.309   &        1.7   & 	    142      \\
  26 	 &  173.690   &     -0.227   &        0.5   & 	     13      \\
  27 	 &  173.701   &     -0.266   &        0.7   & 	     20     \\
  28 	 &  173.725   &     -0.266   &        1.1   & 	     58      \\
  29 	 &  173.690   &     -0.359   &        1.1   & 	     73      \\
  30 	 &  173.662   &     -0.383   &        0.7   & 	     24     \\
  31 	 &  173.305   &      0.033   &        0.9   & 	     35     \\
  32 	 &  173.410   &      0.166   &        0.6   & 	     15      \\
 \enddata  
\end{deluxetable}
 \end{document}